\documentclass[%
 reprint,
superscriptaddress,
 amsmath,amssymb,
 aps,
 floatfix,
 prd,
 nofootinbib,
]{revtex4-2}

\usepackage{graphicx}
\usepackage{dcolumn}
\usepackage{bm}
\usepackage{orcidlink}
\usepackage{xspace}
\usepackage{multirow}
\usepackage{booktabs}
\usepackage{bold-extra}

\newcommand{\Hunit}{\ensuremath{\text{km\,s}^{-1}\,\text{Mpc}^{-1}}}

\newcommand{\Hubble}{\ensuremath{H_{0}}\xspace}

\newcommand{\gwcosmo}{\texttt{gwcosmo}\xspace}

\newcommand{\fullpop}{\textsc{FullPop-4.0}\xspace}
\newcommand{\mltp}{\textsc{MultiPeak}\xspace}

\newcommand{\gladeplus}{\textsc{GLADE+}\xspace}
\newcommand{\gwtc}{GWTC-5.0\xspace}
\newcommand{\params}{\ensuremath{\Lambda}\xspace}

\newcommand{\ndet}{\ensuremath{N_{\text{det}}}\xspace}
\newcommand{\aspire}{\ensuremath{\texttt{ASPIRE}}\xspace}
\newcommand{\dynesty}{\ensuremath{\texttt{dynesty}}\xspace}

\begin{document}

\preprint{APS/123-QED}

\title{Efficient Standard Siren Cosmological Inference with Posterior Reuse}

\author{Alexander Papadopoulos\orcidlink{0009-0006-1882-996X}} \email{a.papadopoulos.1@research.gla.ac.uk}
 \affiliation{Institute for Gravitational Research, University of Glasgow, Glasgow, G12 8QQ, United Kingdom}
 \author{Michael J. Williams\orcidlink{https://orcid.org/0000-0003-2198-2974}}
 \affiliation{Max Planck Institute for Gravitational Physics (Albert Einstein Institute), Am M\"uhlenberg 1, 14476 Potsdam, Germany}
 \affiliation{Institute of Cosmology and Gravitation, University of Portsmouth, Portsmouth PO1 3FX, United Kingdom}
 \author{Christian E. A. Chapman-Bird\orcidlink{https://orcid.org/0000-0002-2728-9612}}
 \affiliation{Institute for Gravitational Wave Astronomy \& School of Physics and Astronomy, University of Birmingham, Edgbaston, Birmingham B15 2TT, United Kingdom}
 \author{Rachel Gray\orcidlink{https://orcid.org/0000-0002-5556-9873}}
\author{Christopher Messenger\orcidlink{https://orcid.org/0000-0001-7488-5022}}
 \affiliation{Institute for Gravitational Research, University of Glasgow, Glasgow, G12 8QQ, United Kingdom}
\date{\today}
\begin{abstract}
Standard siren cosmology depends on several modelling assumptions, including the underlying source population, galaxy catalogue, and detection selection function. Quantifying the effect of each choice typically requires a full, independent analysis. As gravitational-wave catalogs grow incrementally, much computational effort is spent re-analysing largely overlapping datasets. We apply the recently developed \aspire (Accelerated Sequential Posterior Inference via Reuse) sampler to the joint estimation of cosmological and population parameters with dark sirens. \aspire updates existing posterior samples into the posterior under a new hypothesis without restarting from the prior. It combines normalising flows with a sequential Monte Carlo (SMC) scheme to produce accurate posteriors and unbiased evidence estimates. Applying \aspire to a number of dark siren analysis cases with GWTC-5.0, we recover posteriors consistent with nested sampling, while recording a performance improvement of between $6-28\times$ over the set of analyses when using an existing posterior. This enables efficient exploration of modelling choices and rapid updates as catalogs grow, without duplicating existing computational effort.
\end{abstract}

\maketitle

\section{Introduction}\label{sec:intro}
The LIGO-Virgo-KAGRA (LVK) Collaboration~\cite{ligo,virgo,kagra} has detected nearly 400 gravitational-wave (GW) signals from compact binary mergers up to the publication of the Fifth Gravitational Wave Transient Catalog (GWTC-5.0)~\cite{gwtc1,gwtc2,gwtc3,gwtc4results,gwtc5catalog}. These signals contain information which enable investigation into the fundamental nature of the universe. One such avenue is using GW signals to constrain the Hubble constant, $H_0$, first discussed by~\citet{schutz1986} in 1986. When being used for cosmological inference, GW signals are known as \textit{standard sirens}, analogously to standard candles which emit electromagnetic, rather than gravitational, signals \cite{Phillips1993,Riess1995}.
\par
Cosmology with compact binary coalescences (CBCs) requires knowledge of the source luminosity distance, $d_L(z)$, and redshift, $z$~\cite{holz2005,moresco2022}. GW signals directly encode information on the redshifted chirp mass $\mathcal{M}_z = \mathcal{M} (1+z)$, where $\mathcal{M}$ is the source-frame chirp mass, as well as $d_L(z)$~\cite{hughes2002}. This distance depends on redshift through the assumed cosmological model, and given that both this and the detector-frame mass depend on redshift, this degeneracy cannot be broken with GW data alone when considering binary black hole mergers~\cite{messenger2012}. There exists three main classes of standard sirens - bright, dark and spectral. Bright sirens rely on the presence of an electromagnetic counterpart to determine the source redshift~\cite{holz2005,nissanke2010,GW170817,mancarella2024,cozzumbo2025}. When this is not possible, the redshift can be inferred statistically using galaxy surveys~\cite{schutz1986,delpozzo2012,soaressantos2019,Palmese_2020,finke2021,mastrogiovanni2021,gwcosmo2021,icarogw2023,borghi2024}, and by using features in the mass distribution of GW sources~\cite{chernoff1993,taylor2012,farr2019,ezquiaga2021,ezquiaga2022,pierra2026,tagliazucchi2026,bertheas2026} - dark and spectral sirens respectively. 
\par
In practice, these latter two methods are combined to get the most robust constraints on $H_0$. Combining galaxy-catalog information with spectral siren cosmology requires the joint inference of cosmological and compact-binary population parameters within a hierarchical Bayesian framework. Because the CBC mass spectrum has no agreed parametrisation, it can be analysed under a range of models, both weakly and strongly parametrised~\cite{Farah:2024xub,Li:2024rmi,Tong:2025xvd,Gennari:2025nho,Bertheas:2025mzd,Tagliazucchi:2026gxn,Li:2026amt,Gennari:2026dfy,Leyde:2026bat,pierra2026,DeRenzis:2026hcp}. Combined with the many galaxy surveys~\cite{Dalya:2021ewn,McMahon:2026nhi} and alternative tracers~\cite{Beirnaert_2025,lacroze20256} available for dark siren cosmology, and with analyses that incorporate host-galaxy information~\cite{gray2026,agapito2026}, this means the number of possible analyses is ever-expanding. Each of these analyses requires running cosmological inference again from the prior, meaning that the computational cost grows rapidly as more hypotheses are explored. 
\par 
With current detectors, large event localisation volumes and high catalog incompleteness mean that constraints on $H_0$ are dominated by the contribution of spectral sirens; as such, the final results are similar in constraint both with and without galaxy catalog information. Indeed, with \gwtc~\cite{gwtc5catalog,gwtc5_methods} the additional information introduced from dark sirens amounted to less than a $10\%$ improvement in the fiducial $H_0$ constraint compared to spectral sirens~\cite{gwtc5cosmo}. Further to this, ever-growing catalogs of GW events means that hierarchical analyses may be re-run over time using the same modelling assumptions, while adding new events to the analysis. In a sense, the information gained by the previous run is discarded each time the catalog is updated and analysed again from the prior, increasing the total computational time required. Recent works have tried to address this issue in hierarchical analyses, when considering population inference and the neutron star equation of state~\cite{Wolfe:2026dcq,Wouters:2026nbh}. Using variational neural posterior estimation applied to population inference, ~\cite{Wolfe:2026dcq} found that the framework was susceptible to error accumulation when frequently updating with new data. In ~\cite{Wouters:2026nbh}, GPU-accelerated data tempering was used, although the scope remained restricted to chronological event additions rather than broader hypothesis or model modifications.
\par
In the $N$-dimensional space created by the joint inference of cosmological and population parameters, sampling techniques must be used in order to produce $N$-dimensional posterior distributions. This can be done using Markov-Chain Monte-Carlo (MCMC)~\cite{mcmc}, however this does not lend itself to cases where model comparison is desirable, as it does not automatically calculate the Bayesian evidence required to do so. Instead to achieve this, nested sampling is used, which calculates the Bayesian evidence directly while producing the posterior distribution as a by-product~\cite{skilling_proceedings,skilling_paper}. This means that in the case of simple, unimodal likelihoods, MCMC algorithms converge much faster than nested sampling, which must explore the whole prior volume in order to calculate evidences.
\par
In this work we compare posteriors from cosmological inference using nested sampling to posteriors produced with \textit{Accelerated Sequential Posterior Inference via Reuse} (\texttt{ASPIRE})~\cite{Williams:2025aar}, a framework which leverages normalising flows~\cite{Papamakarios:2019fms} and sequential Monte Carlo (SMC)~\cite{fearnhead2010adaptivesequentialmontecarlo,DelMoral2012,Williams:2025szm} to transform existing posterior samples into unbiased posteriors. Using SMC means that rather than restarting analyses from scratch and exploring the full prior, the starting point for the sampler can be an existing posterior. Thus when compared to standard nested sampling algorithms SMC offers the potential for increased performance while still producing the Bayesian evidence required for model comparison.
\par
In this paper, we first present the hierarchical likelihood used in our cosmological analyses and describe the underlying framework of \aspire (\S~\ref{sec:method}). Then in \S~\ref{sec:results} we show the results produced with \aspire for a number of different test cases and compare with posteriors produced by nested sampling. Following this, we demonstrate that \aspire is robust compared to importance sampling in \S~\ref{sec:importance}, then discuss the results and future direction in \S~\ref{sec:dicuss}.

\section{Methodology}\label{sec:method}
We use the hierarchical Bayesian inference pipeline \gwcosmo~\cite{gwcosmo2021,gwcosmo2023,Papadopoulos:2026puy}, which leverages parallel computing via graphics processing units (GPUs), to produce posterior distributions of cosmological and population parameters from GW data and galaxy catalogs.
\subsection{Hierarchical Likelihood}\label{subsec:gwcosmo}
 For a population of GW sources described by hyperparameters \params, with data $\{x_{\text{GW}}\}$ from $\ndet$ sources with detection flag $\{D_{\text{GW}}\}$ denoting a positive detection, the expression for the posterior probability on \params is~\cite{mandel2019,vitale2021,Essick:2023upv}
\begin{equation}\label{eq:hier}
    \begin{aligned}
    p(\params|\{x_{\text{GW}}\},\{D_{\text{GW}}\},I) &\propto p(\params|I) p(\ndet|\params,I) \\
    & \times \prod_{i}^{\ndet} p(x_{\text{GW}i}|D_{\text{GW}i}, \params,I).
    \end{aligned}
\end{equation}
 In this expression, the term $ p(x_{\text{GW}i}|D_{\text{GW}i}, \params,I)$ is the likelihood of a single GW event $i$ given a detection statistic ($D_{\text{GW}i}=1$ indicating detection) and hyperparameters \params. 
 The inclusion of $I$ denotes any further information which is not expressed explicitly. Applying Bayes' theorem and by explicit marginalisation over the source frame parameters of individual GW events $\theta$ (masses, spins, redshift, inclination etc.), we can write this as:
\begin{equation}\label{eq:hier_factorised}
    \begin{aligned}
    p(\params|&\{x_{\text{GW}}\},\{D_{\text{GW}}\},I) 
    \propto p(\params|I)p(\ndet|\params,I) \\ & \times \alpha(\params)^{-\ndet}
   \prod_{i}^{\ndet}\int p(x_{\text{GW}i}|\theta,\params,I)p(\theta|\params,I)d\theta.
    \end{aligned}
\end{equation}
\vspace{5pt}
The term $\alpha(\params)$ is the selection function, given by
\begin{equation}\label{eq:selfunc}
\alpha(\params)=\int p(D_{\text{GW}}|\theta,\params,I)p(\theta|\params,I)d\theta,
\end{equation}
and accounts for the finite detection horizon of GW detectors. This means that the observed population is only ever an incomplete subset of the true source-frame population. This term is the same for each event in the analysis, and so can be removed from the product. While $p(\ndet|\params)$  is seemingly dependent on \params, by using a scale-free Jeffreys prior on the expected rate of detected GW events, the cosmological dependence cancels~\cite{fishbach2018}.
\subsection{Cosmological Inference with \aspire}
The core philosophy behind \aspire is to leverage previous computational work rather than discarding the information produced. Put simply, \aspire can take existing posterior samples and transform them into an unbiased representation of a new target analysis. In order to do this, it follows a two-step process which begins with training a normalising flow, a neural network which creates an invertible mapping between a simple base distribution and the desired distribution, which can be used for sampling or evaluating. This is trained to approximate an initial distribution given some samples (for example, an existing result produced with nested sampling), and this flexible representation of the proposal is then used to inform the start of the analysis. 
\par
In the second step of this process, \aspire uses a series of intermediate target distributions to transform samples from the proposal. Consider a proposal distribution, $\pi_0(\theta)$, approximated by the flow, $q_{\phi}(\theta)$ under the assumption of model $M_1$. Thus, the intermediate distributions as a function of the inverse temperature $\beta_t$ for the updated model $M_2$ are
\begin{equation}
    p_t(\theta|d,M_2,\beta_t) \propto q_{\phi}(\theta)^{1-\beta_t}[p(d|\theta,M_2)p(\theta|M_2)]^{\beta_t},
\end{equation}
where $p(d|\theta,M_2)$ and $p(\theta|M_2)$ are the likelihood and prior for model $M_2$ given data $d$ and parameters $\theta$~\cite{Bernardo2011}. 
The inverse temperature is bound between $\beta_t \in [0,1]$, such that the mutation steps follow $0=\beta_0 < \beta_1 < ... < \beta_T = 1$. This means that at the first and last steps respectively $p(\theta|d,M_2,\beta_t)$ is
\begin{equation}
    \begin{aligned}
    &p_0(\theta|d,M_2,\beta_0) \propto q_{\phi}(\theta), \\
    &p_T(\theta|d,M_2,\beta_T) \propto p(d|\theta,M_2)p(\theta|M_2).
    \end{aligned}
\end{equation}
Here it is clear that by the final annealing step, there remains no information from the proposal distribution in the posterior. The annealing schedule itself is set by solving for the inverse temperature $\beta_t$ that preserves a target effective sample size (ESS)~\cite{Karamanis2022}. The benefit of this is to ensure enough similarity between intermediate temperature steps whilst also maximising computational performance. This means that a higher target ESS will result in a larger number of mutation steps in the sampling process. At each intermediate step, the distribution is resampled using importance weights, derived from the ratio of the current iteration to the previous one, such that $w_t=p_t(\theta|d,M_2,\beta_t)/p_{t-1}(\theta|d,M_2,\beta_{t-1})$. The ESS in this case is defined as the inverse sum over the squared normalised weights, $1/\sum w_t^2$. Multinomial resampling then prevents sample impoverishment by duplicating high-weight particles while discarding lower weights. Finally, in order to decorrelate the distribution and explore the parameter space, the particles are subject to an MCMC step. 
\par
It is also possible for \aspire to use proposal distributions which do not share the same dimensions as the desired posterior. By instead drawing from the prior distribution, the information in the relevant proposal dimensions can still be leveraged. As a natural extension of this, parameters which are shared by the proposal and posterior can also be selected to be drawn from the prior when appropriate. This can in fact increase efficiency of sampling for the geometry of certain problems, for example where correlation structures change significantly or when the proposal does not cover the volume of the posterior.
\par
Overall, this means that \aspire is able to produce unbiased posteriors representing the new target distribution while leveraging previous computational effort. Additionally, it produces estimates of the Bayesian evidence for model comparison, and a required number of posterior samples can be set for the analysis. For a more in-depth treatment of the framework of \aspire, see~\cite{Williams:2025aar}.

\section{Analysis and Results}\label{sec:results}
In this section, we present a number of different applications of \aspire to hierarchical cosmological inference with gravitational waves. This includes:
\begin{itemize}
    \item Adding galaxy catalog information to go from a spectral analysis to dark siren analysis, \textbf{spectral $\to$ dark} (\S~\ref{subsec:spec_to_dark}).
    \item Updating event catalogs chronologically, \textbf{spectral\textsubscript{\textsc{GWTC-4.0}} $\to$ spectral} (\S~\ref{subsec:chronological}).
    \item Using a population-only result as the proposal for cosmological inference, \textbf{pop $\to$ spectral}, and the inverse process, \textbf{spectral $\to$ pop} (\S~\ref{subsec:pop_to_cosmo}).
    \item Changing from one parametrisation of the CBC mass spectrum to another, \textbf{spectral\textsubscript{\mltp} $\to$ spectral} (\S~\ref{subsec:mass_models}).
\end{itemize} 
This order moves from conceptually simplest to most difficult: beginning with targets and proposals which share the same parameter space, and increasing the divergence between them with each analysis. In this section analyses produced with the nested sampler \dynesty~\cite{dynesty_paper,dynesty_v3.1.0}, using its default settings (apart from using $n_\mathrm{live}=1000$ rather than the default $500$), are used as the ground truth, and results produced with \aspire are compared to this, both for posterior agreement and  time comparison. Recovered parameters from both \aspire and \dynesty from the dimensions presented in this section are displayed in Appendix~\ref{app:corner_params}.
Both of the samplers are implemented via a \texttt{bilby}~\cite{bilby_paper} wrapper. Compute times quoted in this work do not include the time required for the pre-processing of data inside \gwcosmo, which typically requires $\mathcal{O}(\text{minutes})$. The compute times also do not include the time required to generate the proposal distribution. This reflects the case in which a proposal already exists, for example from an existing analysis or previous catalog. We consider the total cost, including proposal, in \S~\ref{sec:dicuss}. Additionally, each analysis was repeated for different seeds with identical settings, and the spread of the results from different seeds was found to be below the tolerance used for posterior agreement.
\par 
For all results in this section we use a $4.2$ mnat Jensen-Shannon (JS) divergence threshold for agreement defined in Appendix~\ref{app:js}. The JS divergence is a statistical measure of the difference between two distributions - the greater the JS divergence, the more different they are. Analyses which do not fall below this threshold are considered unsuccessful.
\par
We also include an \aspire spectral siren analysis, which did not use any proposal information, \textbf{prior $\to$ spectral}, for comparison with the other test cases. To reach convergence with out reference posterior, this used a higher target efficiency than others analyses, and as such required significantly more temperature steps than others. This is reflected in the higher runtime and likelihood evaluation count, as shown in Table~\ref{tab:fiducial}, and being only $1.6\times$ faster than the same analysis with \dynesty.

\begin{table*}[htb]
    \centering
    \small
    \setlength{\tabcolsep}{3pt}
    \begin{tabular}{lcrrrrrrrcr}
        \hline
        & \multicolumn{3}{c}{Settings}
        & \multicolumn{2}{c}{$N_{\rm eval}$ ($10^6$)}
        & \multicolumn{3}{c}{Sampling time (hrs)}
        &  $D_\mathrm{JS}^{\mathrm{max}}$ & $\Delta\ln Z$ \\
        \cline{2-4} \cline{5-6} \cline{7-9}
        Analysis
        & Prior$^{\dagger}$ & $N_{\rm steps}$ & $N_{\rm final}$
        & \dynesty & \aspire 
        & \dynesty & \aspire & Speed-up
        & (mnat)
        & \\
        \hline
        \textbf{spectral $\to$ dark}
            & ---          & 20 &  30 & 9.36 & 0.33 & 134.4 &  4.6 & 28.9 & 3.5 & $+0.05 \pm 0.19$ \\
        \textbf{spectral\textsubscript{\textsc{GWTC-4.0}} $\to$ spectral}
            & $\checkmark$ & 20 & 100 & 9.56 & 1.12 & 136.6 & 15.7 &  8.7 & 2.9 & $-0.15 \pm 0.20$ \\
        \textbf{pop $\to$ spectral}
            & $\checkmark$ & 30 &  30 & 9.56 & 1.13 & 136.6 & 16.2 &  8.4 & 1.7 & $-0.08 \pm 0.20$ \\
        \textbf{spectral $\to$ pop}
            & ---          & 30 & 100 & 7.34 & 0.84 & 108.3 & 11.6 &  9.3 & 3.6 & $-0.20 \pm 0.26$ \\
        \textbf{spectral\textsubscript{\mltp} $\to$ spectral}
            & $\checkmark$ & 30 &  50 & 9.56 & 1.47 & 136.6 & 20.5 &  6.6 & 3.7 & $-0.20 \pm 0.20$ \\
        \hline
        \textbf{prior $\to$ spectral}$^{\ddagger}$
            & all          & 30 & 100 & 9.56 & 6.28 & 136.6 & 87.0 &  1.6 & 3.0 & $+0.03 \pm 0.19$ \\
        \hline
    \end{tabular}
    \caption{Fiducial settings and results for the analyses presented in this work.
    The variables $N_\mathrm{steps}$ and $N_\mathrm{final}$ represent the number of MCMC steps per mutation step and the final number of MCMC steps for resampling respectively.
    In every case we used $N_{\rm particles}=2500$ particles, required $N_{\rm samples, final}=5000$
    final samples, and used a target efficiency of $[0.5,\,0.95]$. The initial results
    were produced by \dynesty\ with 1000 live points.
    $^{\dagger}$Analyses marked $\checkmark$ draw $\lambda_g$, $\lambda_{g,\rm low}$, $\mu_{g,\rm low}$ and $\sigma_{g,\rm low}$ from the prior.
    the total number of likelihood evaluations is denoted by $N_{\rm eval}$; times are the \texttt{sampling\_time} recorded by each run on a single Nvidia H100 GPU, and the speed-up is their ratio.
    The bootstrap noise floor of $0.0042$~nat is compared with $D_\mathrm{JS}^{\mathrm{max}}$, the maximum over the shared parameters of the marginal Jensen--Shannon divergence against the \dynesty\ reference.
    The difference in evidences is calculated as $\Delta\ln Z = \ln Z_{\aspire} - \ln Z_{\dynesty}$, with the two uncertainties added in quadrature; every analysis is consistent with zero. $^{\ddagger}$The \textbf{prior $\to$ spectral} analysis did not include any proposal information, with all parameters being drawn from the prior, and in this case the target efficiency was allowed to vary between $[0.9,\,0.99]$, resulting in more temperature steps than in other cases.}
    \label{tab:fiducial}
\end{table*}

\subsection{Data}\label{subsec:data}
In the following sections, we make use of publicly available data from the Fifth Gravitational-Wave Transient Catalog (\gwtc)~\cite{gwtc5_pe_part2,gwtc5_pe_part1,gwtc5catalog}. This includes posterior samples and skymaps for GW events, as well as injection campaigns to quantify selection effects~\cite{Essick:2025}. The full set of GW events used in our analyses are chosen using a selection threshold of false alarm rate (FAR) $<0.25 \,\mathrm{yr^{-1}}$ which yields 235 standard sirens, excluding the bright siren GW170817~\cite{GW170817,gwtc5cosmo}. For dark siren analyses we use the \gladeplus galaxy catalog with luminosity weighting~\cite{Dalya:2021ewn}. Unless otherwise indicated, we model the mass distribution of GW sources using the \fullpop mass model~\cite{gwtc4rp,gwtc4cosmo}, following the fiducial analyses in~\cite{gwtc5cosmo}. In all cases, we parametrise the redshift distribution using a Madau-Dickinson distribution~\cite{mddist}. For a description of the parameters highlighted in this work, as well as the prior ranges used, see App.~\ref{app:corner_params}.

\subsection{Adding Galaxy Catalog Information}\label{subsec:spec_to_dark}
Current dark siren analyses find only a minor improvement in \Hubble constraint (less than $10\%$) relative to spectral-only analyses. Since these analyses all contain the same spectral information, the similarity of their constraints indicates that they are repeatedly processing largely the same information, with only the galaxy catalog and weighting-model contribution changing. Each analysis which begins from the prior includes the same spectral information while injecting new information from a different survey or weighting model. Rather than effectively discarding the spectral contribution each time, using \aspire means that the spectral result can be rapidly updated to include information from a chosen galaxy catalog.
\begin{figure}[h]
    \includegraphics[width=\linewidth]{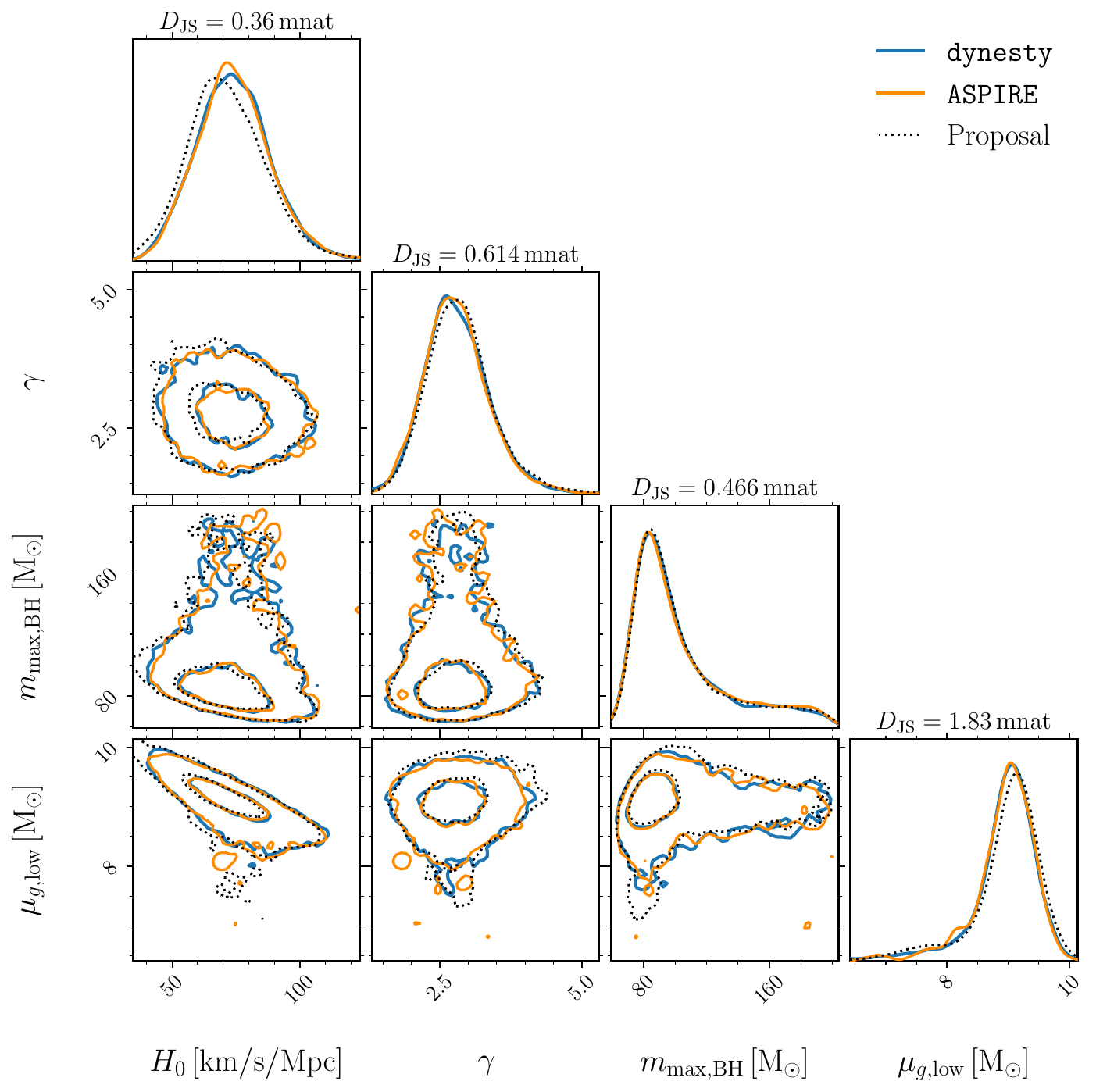}
    \caption{Posterior distributions for a subset of \fullpop parameters, updating a spectral-only analysis with galaxy catalog information (\textbf{spectral $\to$ dark}). The curves are the spectral proposal (black dotted line), dark siren \dynesty reference (blue solid line) and dark siren \aspire result (orange solid line). The contours shown in the 2-D joint posteriors are at the $68\%$ and $90\%$ confidence intervals respectively. These were produced using \gwtc data and the \gladeplus galaxy catalog with luminosity weights. For this analysis, all parameters fall beneath the $4.2$ mnat JS divergence threshold, with the per-parameter values being shown above the 1D plots.}
    \label{fig:spectral_dark}
\end{figure}
\par
The posteriors shown in Fig.~\ref{fig:spectral_dark} show good agreement between the reference \dynesty result and the result produced with \aspire. This is confirmed by the fact that all 23 parameters fall beneath the Jensen-Shannon threshold for convergence. When comparing total sampling time in Table~\ref{tab:fiducial}, \aspire is significantly faster than \dynesty, with the former requiring 4.6 hours, two orders of magnitude less than the 134.4 hours required by the latter. Indeed, this is also reflected by the fact that \aspire requires an order of magnitude fewer likelihood evaluations to reach convergence. Additionally, we see here that the evidences from the two methods are in agreement within the \dynesty errors. This is not a surprising result, given that the proposal and target distributions for this analysis overlap almost entirely, differing only marginally in \Hubble and its correlated parameters. 
\subsection{Updating Event Catalogs Over Time}\label{subsec:chronological}
A natural usage for \aspire in the hierarchical context is to take an existing result from a previous catalog of GW sources and add more information in the form of additional, newly-observed events. Indeed, assuming that the modelling assumptions remain the same (catalog, mass model etc), the posterior obtained from additional events should fall entirely within the support of the proposal. A similar analysis was previously performed in~\cite{Wolfe:2026dcq}, which used neural Bayesian updating, and considered incremental updates to the catalog, as well as the whole updated catalog, but without inferring cosmology. To test this in our framework, we run a spectral \dynesty analysis on 141 GWTC-4.0 events and with consistent sensitivity injections (a selection which is consistent with the analyses presented in~\cite{gwtc4cosmo}) before running an identical \aspire analysis with GWTC-5.0 events (both using the same selection cut in this case, as mentioned in \S~\ref{subsec:data}).
\begin{figure}[h]
    \includegraphics[width=\linewidth]{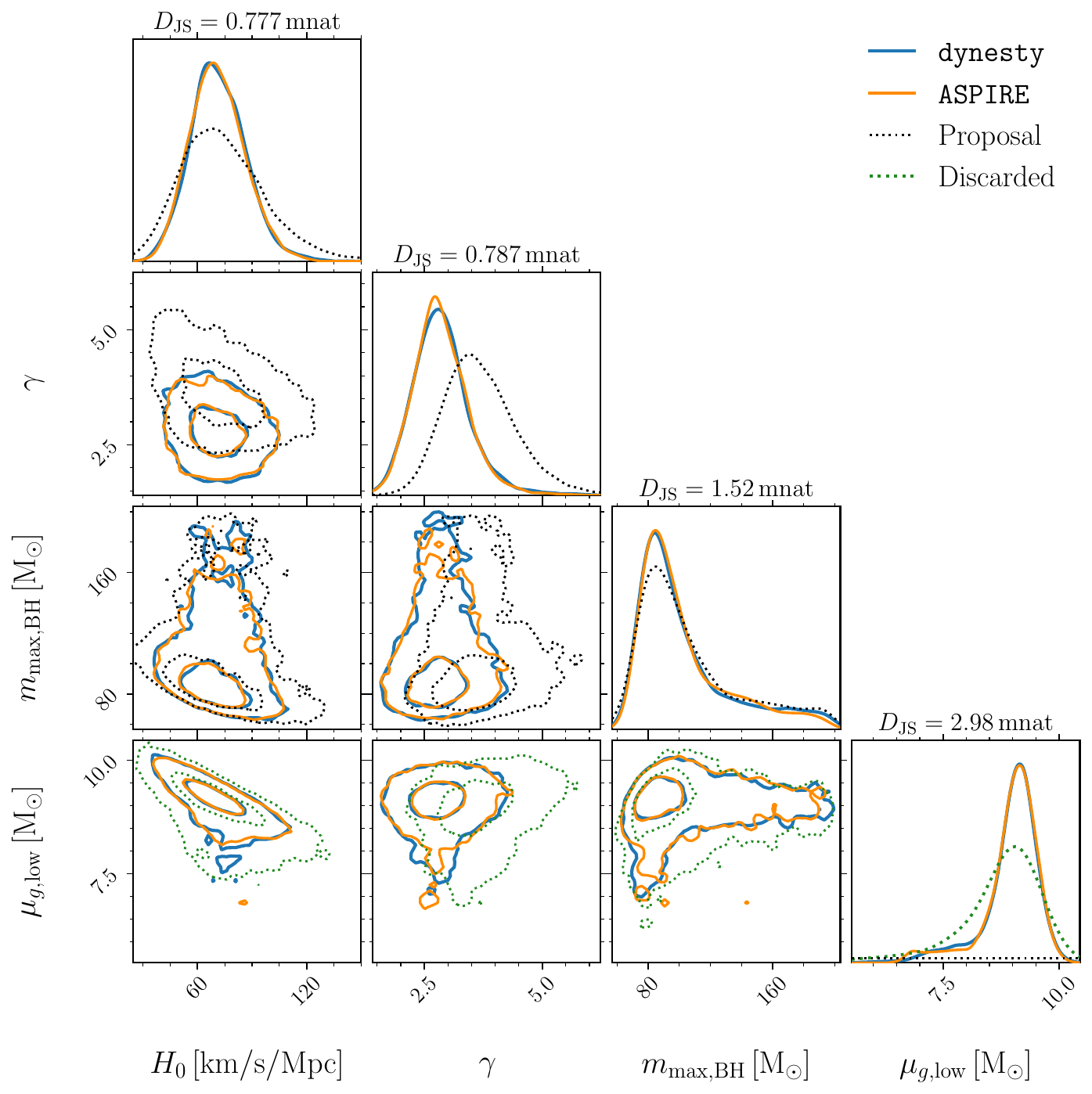}
    \caption{Posterior distributions for a subset of parameters when updating event catalogs over time (\textbf{spectral\textsubscript{\textsc{GWTC-4.0}} $\to$ spectral}), with the spectral proposal (black dotted line), \dynesty reference (blue solid line) and \aspire result (orange solid line). Parameters which were discarded from the proposal as well as their 2D regions are shown with a green dotted line, and were instead sampled from a uniform prior. The contours shown in the 2-D joint posteriors are at the $68\%$ and $90\%$ confidence intervals respectively. These posteriors were produced using \gwtc data and include no galaxy catalog information. For this analysis, all parameters fall beneath the $4.2$ mnat JS divergence threshold, with the per-parameter values being shown above the 1D plots.}
    \label{fig:chronological}
\end{figure}
\par
Updating our GW catalog with \aspire required only 15.7 hours, compared with 136.6 hours for the same \dynesty analysis run from the prior - a factor of almost $9\times$ improvement in performance by utilising posterior reuse. In this case, in order to meet the threshold for posterior agreement in a competitive timescale, four parameters from the proposal had to be discarded and instead drawn from the prior - $\lambda_g,\ \lambda_{g,\rm low},\ \mu_{g,\rm low},\ \sigma_{g,\rm low}$ - describing the most cosmologically informative feature, the low-mass peak in the BBH mass distribution. This is partially shown in Fig.~\ref{fig:chronological}, where the discarded parameter is denoted with a green dotted line, and the uniform prior is instead shown with a black dotted line. This was required not because of a change to the parameter space - both the target and proposal use identical modelling assumptions - but because the updated dataset changes the correlations inside the posterior. Indeed, for example the correlation between $\lambda_g$ and $\mu_{g,\rm low}$ changes from $-0.39$ to $-0.66$ from the proposal to the target posterior. This is despite the fact that the proposal is benign in other ways - containing the bulk of the target posterior in its width and being centred in the same location. Due to this rotation, it becomes more efficient for the sampler to start with the broader, prior-drawn proposal than to try and use the complex and correlated proposal. 
\subsection{Turning Cosmological Inference On (or Off)}\label{subsec:pop_to_cosmo}
Analyses which seek to understand the structure of the compact binary mass distribution are prevalent given the ever-expanding catalog of GW sources. These analyses choose a fiducial cosmology and fix this, and do not include information from galaxy surveys when inferring constraints on the mass spectrum. Cosmological inference with GWs requires the addition of only one degree of freedom, given that spectral (and dark) siren analyses must jointly infer the shape of the compact binary mass spectrum in order to provide robust results. Rather than duplicate the inference of the mass spectrum parameters, it is possible with \aspire to use existing information to complete cosmological inference efficiently from population-only posteriors, or to do the inverse, and use an existing dark siren analysis to inform us about just the population.
\par
\begin{figure}[h]
    \includegraphics[width=\linewidth]{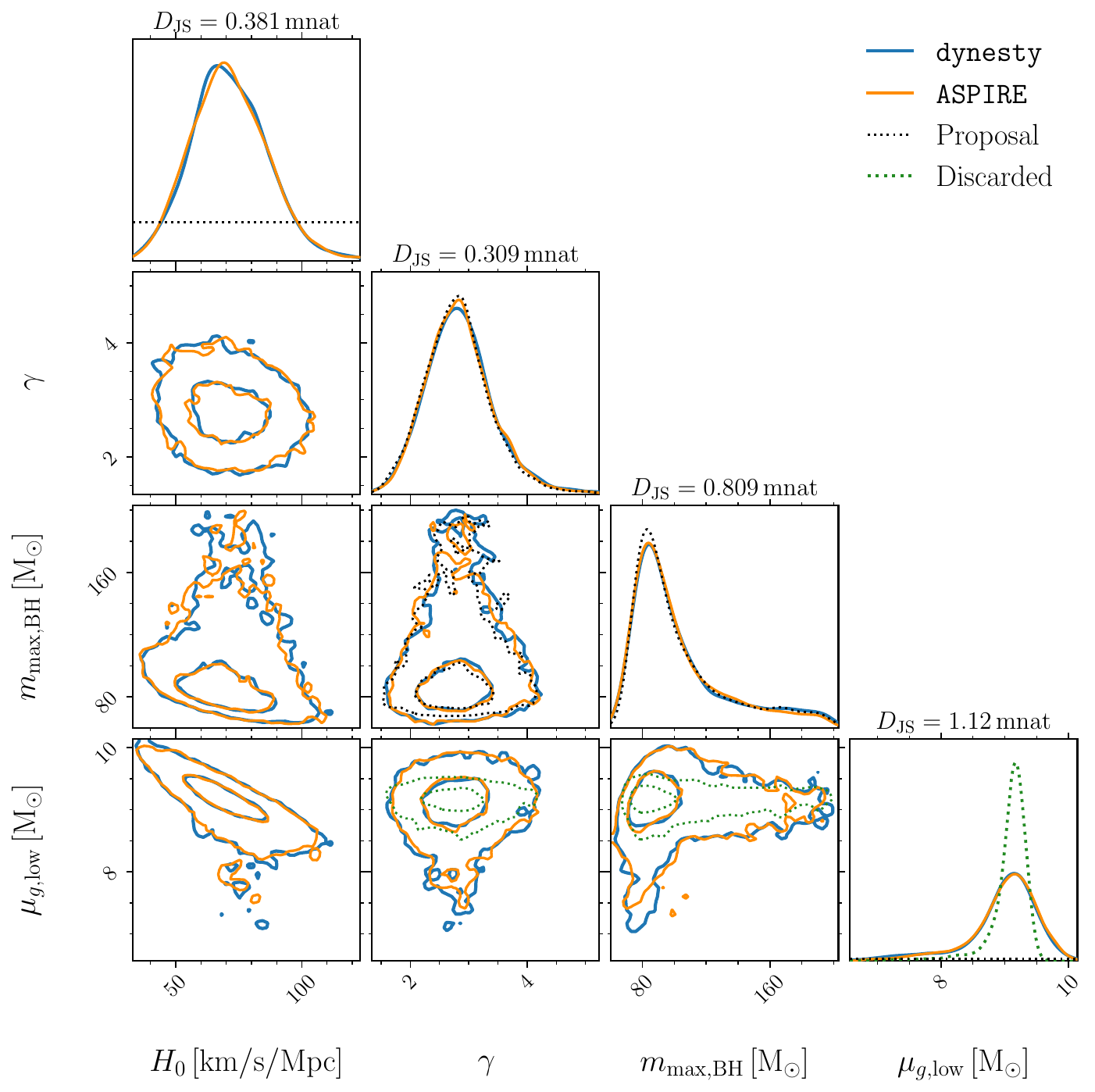}
    \caption{Posterior distributions for a subset of parameters, adding a degree of freedom to a population-only analysis by varying cosmology (\textbf{pop $\to$ spectral}). The curves include the population-only proposal (black dotted line), \dynesty reference (blue solid line) and \aspire result (orange solid line). Parameters which were discarded from the proposal as well as their 2D regions are shown with a green dotted line, and were instead sampled from a uniform prior. In this analysis, \Hubble was also drawn from a uniform prior as this dimension was not included in the proposal. The contours shown in the 2-D joint posteriors are at the $68\%$ and $90\%$ confidence intervals respectively. These posteriors were produced using \gwtc data and includes no galaxy catalog information. For this analysis, all parameters fall beneath the $4.2$ mnat JS divergence threshold, with the per-parameter values being shown above the 1D plots.}
    \label{fig:from_pop}
\end{figure}

These two tests have two inverse challenges, the first (\textbf{pop $\to$ spectral}) requires a larger number of mutation steps, as shown in Table~\ref{tab:fiducial}, due to the proposal having a smaller width than the target posterior. Additionally, the same four parameters as \S~\ref{subsec:chronological} are drawn from the prior, further widening the proposal. Given that these parameters, $\lambda_g,\ \lambda_{g,\rm low},\ \mu_{g,\rm low},\ \sigma_{g,\rm low}$, correlate with \Hubble, they see the greatest divergence between target and proposal. An example of this can be seen in Fig.~\ref{fig:from_pop}, shown by the green discarded posterior on $\mu_{g,\rm low}$. These additional stages lengthen the sampling time, while still being much faster than \dynesty, with a performance improvement of $8.4\times$. 
\par
\begin{figure}[h]
    \includegraphics[width=\linewidth]{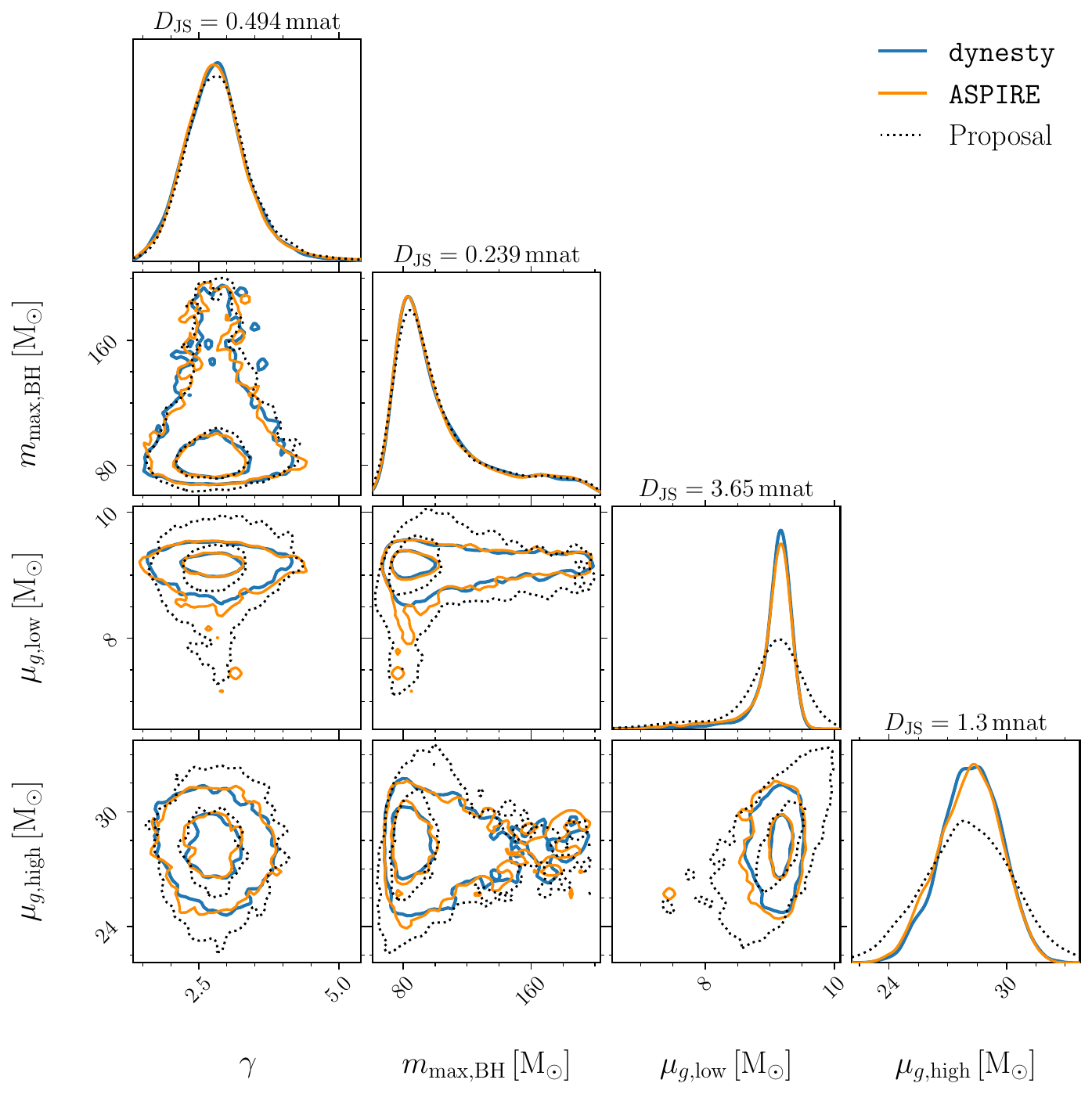}
    \caption{Posterior distributions for a subset of parameters, transforming a spectral siren analysis with varying \Hubble to a fixed-\Hubble result (\textbf{spectral $\to$ pop}). The curves are the spectral proposal (black dotted line), population-only \dynesty reference (blue solid line) and corresponding \aspire result (orange solid line). The contours shown in the 2-D joint posteriors are at the $68\%$ and $90\%$ confidence intervals respectively. These posteriors were produced using \gwtc data and include no galaxy catalog information. For this analysis, all parameters fall beneath the $4.2$ mnat JS divergence threshold, with the per-parameter values being shown above the 1D plots.}
    \label{fig:to_pop}
\end{figure}

Using a spectral siren result as proposal for a fixed cosmology analysis (\textbf{spectral $\to$ pop}, shown in Fig.~\ref{fig:to_pop}) is conceptually easier - the fixed cosmology posteriors are in a sense a conditional distribution at a fixed value of \Hubble (in this case $\Hubble=70\,\Hunit$), and as such should be contained within the width of the proposal. This proves to be true, as seen in Table~\ref{tab:fiducial} this analysis shows a performance improvement against \dynesty of $9.3\times$, and did not require any parameters to be drawn from the prior. This confirms that with \aspire, posteriors contained within the proposal that require only a reduction in width are more straightforward to produce, as they are already fully supported by the proposal. 

\subsection{Changing Mass Model}\label{subsec:mass_models}
Given the uncertainty over the exact form of the CBC mass distribution, a number of different and increasingly complex mass distributions can be used in analyses. For example, the \textsc{MultiPeak} model used in~\cite{gwtc4cosmo,gwtc5cosmo} (which covers only the black hole mass range) naturally extends to the \fullpop model when also modelling the neutron star mass range. Rather than discarding the information contained in the black-hole only analysis, this can be used as the proposal for a full-CBC spectrum analysis.
\par
\begin{figure}[h]
    \includegraphics[width=\linewidth]{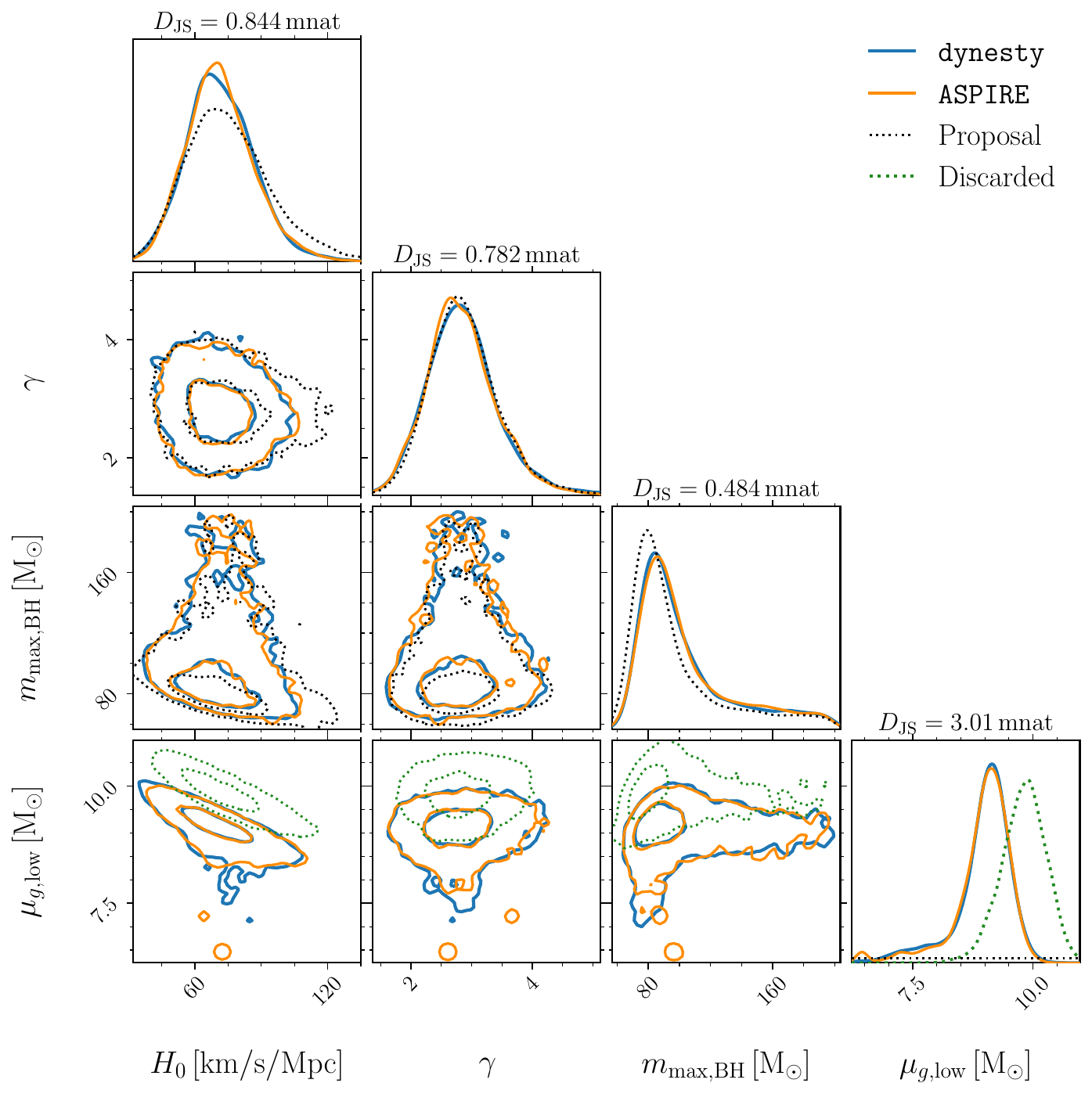}
    \caption{Posterior distributions for a subset of parameters, with the spectral proposal (black dotted line), \dynesty reference (blue solid line) and \aspire result (orange solid line), for an analysis changing from one mass model to another (\textbf{spectral\textsubscript{\mltp} $\to$ spectral}). Parameters which were discarded from the proposal as well as their 2D regions are shown with a green dotted line, and were instead sampled from a uniform prior. The contours shown in the 2-D joint posteriors are at the $68\%$ and $90\%$ confidence intervals respectively. These posteriors were produced using \gwtc data and include no galaxy catalog information. For this analysis, all parameters fall beneath the $4.2$ mnat JS divergence threshold, with the per-parameter values being shown above the 1D plots.}
    \label{fig:mass_models}
\end{figure}

In this case, a spectral \textsc{MultiPeak} analysis is used as the proposal distribution for a spectral \fullpop analysis. However, for parameters which are not present in the \textsc{MultiPeak} model, primarily those describing the neutron star mass range, sampling for the proposal is done directly from the uniform prior. This analysis proves to be the most challenging and computationally expensive, as seen  in Table~\ref{tab:fiducial}, seeing a performance improvement of $6.6\times$ over the reference nested sampling result. The reason for this is quite clear, as the majority of parameters in this run (12/23) have to be sampled from the prior, diminishing the information provided by the proposal distribution. Eight of these are additional parameters of the \fullpop relative to the \textsc{MultiPeak} and have to be drawn in this way. The remaining prior-drawn parameters are common to the analyses in \S~\ref{subsec:chronological} and \S~\ref{subsec:pop_to_cosmo}. Shown in Fig.~\ref{fig:mass_models}, the cosmologically informative parameter $\mu_\mathrm{g,low}$ which correlates with \Hubble is shifted in the proposal relative to the target distribution. The cause of this shift is difficult to determine a-priori, as the proposal model uses only BBH events, while the target posterior uses all CBCs above the selection threshold. Irrespective of the cause, this makes producing unbiased posteriors on $\mu_\mathrm{g,low}$ and it's correlated parameters computationally challenging, and so it becomes more efficient to simply sample from the uniform prior distribution. 

\section{Comparison with importance sampling}\label{sec:importance}
In order to highlight the strengths of this method, we also compare it to the results produced by utilising importance sampling (IS). This is done in two ways - first by directly reweighting the existing proposal samples, using uniform priors in the cases where the model dimensions change between proposal and target.
Importance sampling estimates a target distribution $p(\theta)$ using samples drawn from a different proposal distribution $q(\theta)$. Each sample $\theta_i \sim q$ is assigned a weight $w_i \propto \mathcal{L}(\theta_i)\,\pi(\theta_i)/q(\theta_i)$, normalised so that $\sum_i w_i = 1$, which requires one likelihood evaluation per sample. The method is efficient only when $q$ closely resembles $p$, where they differ, a small number of samples dominate the weights and the effective number of independent samples collapses. As a stopping criterion we require each test to achieve an ESS of 5000, or failing this to complete $N_{\text{eval}}=2\times10^6$ likelihood evaluations, which is a  greater number than any of the \aspire tests.
\par
As can be seen in Table~\ref{tab:is}, this is effective for the case in \S~\ref{subsec:spec_to_dark} as the posteriors show a high degree of similarity, and so the efficiency, $\epsilon$, remains very high and the desired effective sample size (ESS) of $5000$ is reached while also passing the JS threshold in much less time than \aspire. Here, the ESS is again defined as $1/\sum w_i^2$, or more clearly the expected number of samples if all sample weights were equal, while $\epsilon$ is defined as the ESS per number of likelihood evaluations, $\frac{\mathrm{ESS}}{N_{\mathrm{eval}}}$. In all other cases, the proposal and target distributions are too dissimilar, which significantly degrades $\epsilon$ meaning that it becomes impossible to reach the desired ESS with the number of samples available. 
\par
The second comparison takes the flow fitted to the proposal by \aspire, and can be used to draw as many samples as are required. In this case, a more direct comparison is made, as the parameters which were discarded from the \aspire proposal are also sampled from the prior. Despite this, it is still clear that for the majority of cases, importance sampling is not preferable to using \aspire. Other than \textbf{spectral $\to$ dark} \S~\ref{subsec:spec_to_dark}, where IS takes approximately the same time as SMC whilst still passing the JS divergence threshold, no other case can successfully replicate the output or performance of \aspire. Indeed, although converting a cosmology analysis into a population-only one took less than half of the time required by \aspire to reach the required $\mathrm{ESS}=5000$, it did not pass the threshold required for agreement, as despite reaching the stopping criterion its ESS was dominated by a small number of samples. The remaining analyses hit the budget of $2\times10^6$ evaluations, without reaching the required ESS or agreement thresholds, showing the unsuitability of importance sampling to these cases.
\par 
In Appendix~\ref{app:flow} we examine the effect of different flow settings on the fitted distribution by sampling from it and comparing to a true posterior. Over a set of different available flow configurations in this framework, we find that the default settings used here provide the best fit to the true posterior. This allows us to rule out this lever as affecting the results of either the SMC or especially the importance sampling from \aspire's flow. 
\begin{table*}[htb]
    \centering
    \begin{tabular}{llrrrrcc}
        \hline
        Analysis & Method & $\epsilon$ (\%)
        & $N_\mathrm{eval}$ & ESS & $t$ (hrs)
        & $D_\mathrm{JS,max}$ & Pass \\
        \hline
        \multirow{3}{*}{\textbf{spectral $\to$ dark}} & \aspire\ (SMC)
          &  1.515 &   330\,000 & 5000.0 &  4.65 & 0.0035 & \checkmark \\
        & IS, source samples & 89.91 &     5\,562 & 5000.0 & 0.09 & 0.0029 & \checkmark \\
        & IS, \aspire\ flow &  1.468 &   340\,538 & 5000.0 &  4.63 & 0.0032 & \checkmark \\
        \hline
        \multirow{3}{*}{\textbf{spectral\textsubscript{\textsc{GWTC-4.0}} $\to$ spectral}} & \aspire\ (SMC)
          &  0.445 & 1\,122\,500 & 5000.0 & 15.73 & 0.0029 & \checkmark \\
        & IS, source samples &  0.946 &     6\,617 &   62.6 &  0.09 & 0.0252 & $\times$ \\
        & IS, \aspire\ flow & 0.00025 & 2\,000\,000 &    5.1 & 23.61 & 0.174 & $\times$ \\
        \hline
        \multirow{3}{*}{\textbf{pop $\to$ spectral}} & \aspire\ (SMC)
          &  0.443 & 1\,127\,500 & 5000.0 & 16.20 & 0.0017 & \checkmark \\
        & IS, source samples &  3.974 &     6\,668 &  265.0 &  0.09 & 0.107 & $\times$ \\
        & IS, \aspire\ flow & 0.00027 & 2\,000\,000 &    5.3 & 24.69 & 0.208 & $\times$ \\
        \hline
        \multirow{3}{*}{\textbf{spectral $\to$ pop}} & \aspire\ (SMC)
          &  0.599 &   835\,000 & 5000.0 & 11.61 & 0.0036 & \checkmark \\
        & IS, source samples &  0.257 &     6\,999 &   18.0 &  0.10 & 0.0394 & $\times$ \\
        & IS, \aspire\ flow &  1.639 &   304\,978 & 5000.0 &  4.15 & 0.0139 & $\times$ \\
        \hline
        \multirow{3}{*}{\textbf{spectral\textsubscript{\mltp} $\to$ spectral}} & \aspire\ (SMC)
          &  0.341 & 1\,467\,500 & 5000.0 & 20.50 & 0.0037 & \checkmark \\
        & IS, source samples &  0.019 &     6\,774 &    1.3 &  0.10 & 0.657 & $\times$ \\
        & IS, \aspire\ flow & 0.00011 & 2\,000\,000 &    2.2 & 24.62 & 0.648 & $\times$ \\
        \hline
    \end{tabular}
    \caption{Importance sampling against \aspire. $\epsilon = \mathrm{ESS}/N_\mathrm{eval}$ is
    effective samples per target-likelihood call, with ESS being defined as $1/\sum w_i^2$, inverse of the sum over the squared sample weights. For all cases of importance sampling, the sampling ran for either as many evaluations as were required to reach an ESS of $5000$, to compare with \aspire, or a maximum of $2\times10^6$ evaluations to avoid unnecessary computational cost. In the case of the sample-based IS test, all the available proposal samples were used unless the ESS threshold was reached. The quoted time taken includes only the run time required for sampling, and the JS divergence value is calculated similarly to other places in this work, being the maximum over the shared parameters of the marginal JS divergence. The final column shows whether or not the run passes the noise threshold of $0.0042$ nat established in App.~\ref{app:js}.}
    \label{tab:is}
\end{table*}

\section{Discussion and outlook}\label{sec:dicuss}

In this work, we have demonstrated a novel method of producing unbiased posteriors on cosmological and population parameters from standard sirens using \aspire. This framework had previously been shown to be applicable to the case of single-GW event parameter estimation~\cite{Williams:2025aar}, and here it is shown to be applicable in a hierarchical context as well. Rather than beginning each analysis from the prior, here we have shown that it is possible to leverage existing computational effort to analyse a number of modelling hypotheses in far less time than would be required by field-standard nested sampling.
\par
The framework of SMC with posterior reuse improves on the performance of \dynesty by a factor of at least $6.6\times$ across our suite of fiducial analyses when using a proposal distribution. The effect of the additional proposal information which \aspire uses to optimise the sampling process is reflected by the fact that it requires an order of magnitude fewer likelihood evaluations than \dynesty, as shown in Table~\ref{tab:fiducial}. In particular, the \textbf{prior $\to$ spectral} analysis being slower than other \aspire runs demonstrates that the additional proposal information significantly improves the sampler's performance. This takes convergence times from multiple days to less than a day in every case presented in this work when utilising posterior reuse,  while still meeting our threshold for posterior agreement. In the course of this study we have highlighted some of the strengths and weaknesses in the applicability of this framework. Adding catalog information to a spectral-only posterior (\S~\ref{subsec:spec_to_dark}) showed the most improved performance in our analyses, by a factor of $28.9\times$ compared to \dynesty. This analysis provided an easy test case for \aspire, as the two sets of posteriors overlap almost completely, with the exception of \Hubble and the handful of parameters which correlate most strongly with it. This may not be the case in the future however, as improved galaxy catalogs or other tracers are incorporated into dark siren analyses and provide more information~\cite{borghi2024}. 
\par
\begin{figure}[h]
    \includegraphics[width=\linewidth]{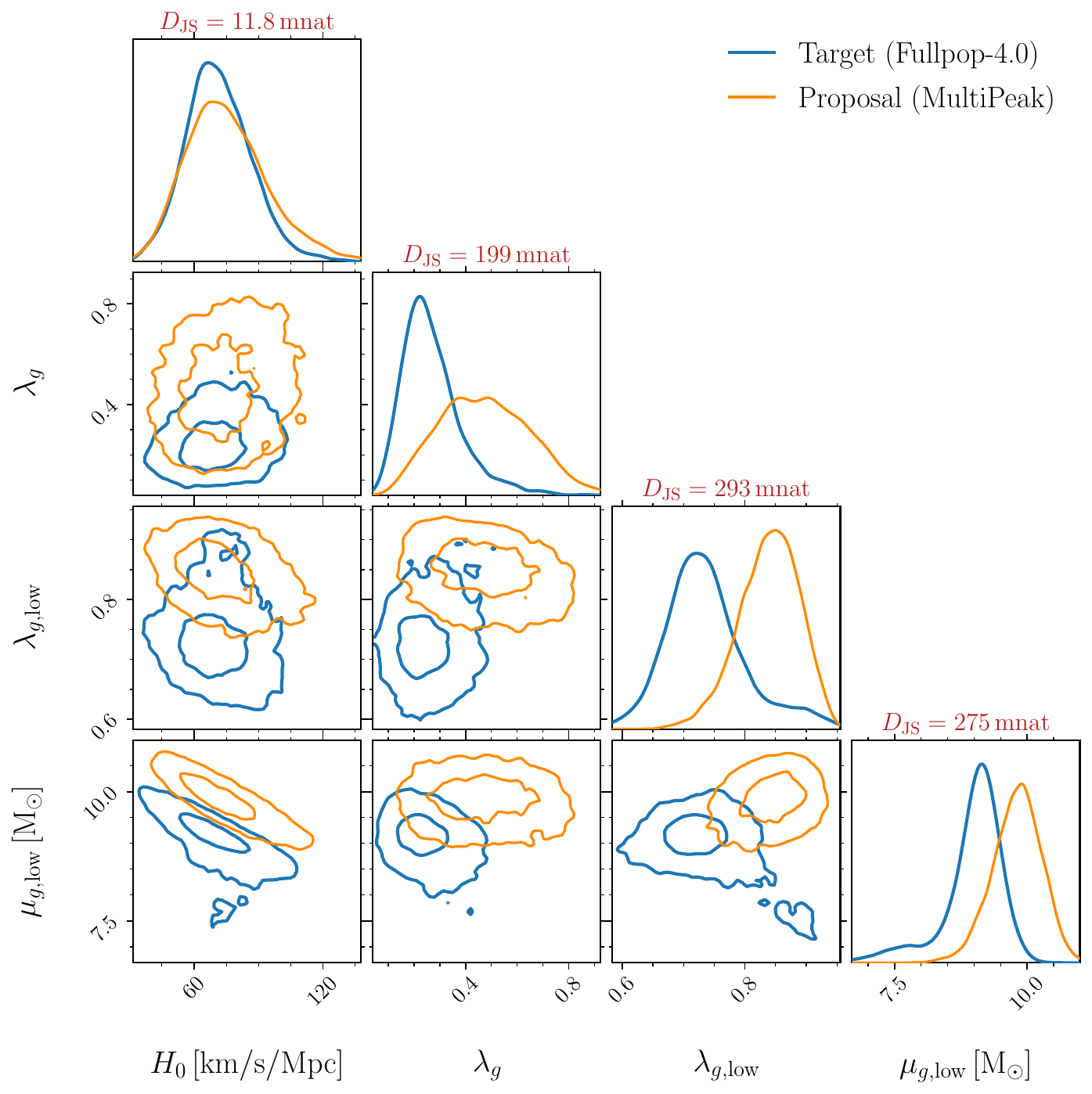}
    \caption{Parameters sampled from the prior rather than proposal distribution when changing mass models (\S~\ref{subsec:mass_models}) and \Hubble. The difference between the target (blue) and proposal (orange) is stark, with several parameters only showing a slight degree of overlap in the joint space.}
    \label{fig:mass_models_params}
\end{figure}

A more challenging application is shown in updating a previous GW event catalog with a new one (\S~\ref{subsec:chronological}), or using a fixed-cosmology run as a proposal for a spectral siren analysis (\S~\ref{subsec:pop_to_cosmo}). Both of these analyses demonstrated significant performance improvements, $8.7\times$ and $8.4\times$ respectively, but faced different issues. Updating GWTC-4.0 using \gwtc showed that changing correlations inside the posterior made \aspire less efficient, requiring a much larger number of mutation steps in the temperature schedule. The fixed cosmology proposal analysis was less efficient than more straightforward analyses, but this time was caused by a weakness of this method - SMC with posterior reuse struggles to broaden an already-tight proposal. This is inherent, as the reweighting and resampling steps of the process cannot create samples where none exist in the proposal, and so only the expensive mutation steps can disperse the particles. As shown in our analyses, this can be resolved by using the prior, or a broader distribution as the proposal. For both of the above cases, the efficiency was found to improve when sampling a subset of parameters from a uniform prior rather than the proposal. The inverse of the fixed-cosmology analysis, removing the cosmological dimension, shows that the opposite case where parameters narrow while being centred on the same location, is a straightforward application of SMC with posterior reuse. Lastly, changing mass models was shown to provide another distinct test of the framework, with a larger dataset and more parameters providing more complex posteriors to be explored. This is illustrated in Fig.~\ref{fig:mass_models_params} which shows the correlations between the affected parameters. This was ameliorated as before by sampling the parameters directly from the prior. As with all the results in this work, optimising the analysis settings was easier given that a known ground truth was available for comparison. As noted in~\cite{Williams:2025aar}, knowing a priori what settings are required to ensure convergence is an open question. More conservative settings ensure that the analysis is robust while sacrificing performance (more mutation and MCMC steps, meaning more evaluations).
\par
The speed-up factors in Table~\ref{tab:fiducial} assume that the proposal distribution already exists, and does not include the cost of producing it. If instead a suite of analyses are produced from scratch the cost of this proposal must be included. In our tests, the \textbf{spectral} result can serve as a hub from which to produce other analyses. For example, producing the \textbf{spectral}, \textbf{dark} and \textbf{pop} posteriors independently with \dynesty requires $136.6 + 134.4 + 108.3 = 379.3$ hours. Using \aspire with no proposal for the \textbf{spectral} analysis, followed by the \textbf{spectral $\to$ dark} and \textbf{spectral $\to$ pop} updates, requires $87.0 + 4.6 + 11.6 = 103.2$ hours, a $3.7\times$ improvement over \dynesty\footnote{This assumes that an \aspire proposal serves at least as well as the \dynesty one; since both pass the same agreement threshold we expect this to hold.}.  The same framework but with \dynesty producing the proposal yields a total time of $152.8$ hours, or a $2.5\times$ speed-up. The overall gain increases with each analysis run with \aspire, as the initial cost of generating the proposal is paid once, and each additional analysis costs $\mathcal{O}(10)$ hours.
\par
\begin{table*}
    \centering
    \small
    \setlength{\tabcolsep}{6pt}
    \begin{tabular}{lccc}
        \hline
        Analysis & $D_\mathrm{KL,proposal}$ [nat] & $D_\mathrm{KL,prior}$ [nat] & Fraction \\
        \hline
        \textbf{spectral $\to$ dark} & $ 1.19 \pm 0.04$ & $27.33 \pm 0.03$ & $4.4\%$  \\
        \textbf{spectral\textsubscript{\textsc{GWTC-4.0}} $\to$ spectral}       & $12.78 \pm 0.08$ & $26.46 \pm 0.03$ & $48.3\%$ \\
        \textbf{pop $\to$ spectral}     & $14.40 \pm 0.10$ & $26.46 \pm 0.03$ & $54.4\%$ \\
        \textbf{spectral $\to$ pop}     & $ 1.73 \pm 0.04$ & $25.95 \pm 0.03$ & $6.7\%$  \\
        \textbf{spectral\textsubscript{\mltp} $\to$ spectral}         & $21.72 \pm 0.09$ & $26.46 \pm 0.03$ & $82.1\%$ \\
        \hline
    \end{tabular}
    \caption{Comparison of information gained by \aspire between the proposal distribution or prior, and posterior, using the KL divergence to quantify the information. In all cases, using the \aspire proposal distribution meant that the sampler had to gain less information to reach the posterior.}
    \label{tab:proposal_kl}
\end{table*}
As shown in Table ~\ref{tab:proposal_kl}, the information gained by \aspire differs when comparing runs from the prior and proposal distributions. Here we use the Kullback-Leibler (KL) divergence~\cite{kldiv} to quantify information gain, which unlike the JS divergence is not symmetric. The information gained between proposal (which includes parameters sampled from the prior) and posterior follows a similar pattern to the total sampling time and likelihood evaluations shown in Table~\ref{tab:fiducial}. Th higher percentages correspond to the analyses which had parameters drawn from the prior, so that less information was added by the proposal. Analyses which require less information to reach the posterior correspondingly require fewer likelihood evaluations, and thereby less total sampling time. This demonstrates the benefit of the information added by the proposal, leaving the sampler with less information to gain than otherwise.
\par
We also compared the usage of \aspire to two importance-sampling tests, intended to determine if it was possible to complete the same analyses even more efficiently by avoiding the overhead associated with \aspire, the results of which are shown in Table~\ref{tab:is}. In the first test, importance sampling from only the available posterior samples, the \textbf{spectral $\to$ dark} analysis was able to reproduce the target posterior in a matter of minutes, significantly outperforming \aspire. While not surprising due to the degree of overlap in the target and proposal posteriors, this may not be replicated if using a more informative galaxy catalog in future analyses. In every other sample-based test, importance sampling was not able to generate a large enough ESS to be comparable. The second test, importance sampling from the same flow as the SMC runs, yielded similar results, with only the \textbf{spectral $\to$ dark} test passing our agreement threshold and with approximately the same performance as the equivalent \aspire test. Of interest is the flow based importance sampling of the \textbf{spectral $\to$ pop} test, which achieved the required ESS of $5000$, but failed the JS divergence criterion as a number of parameters were still over-dispersed relative to the target. In the remaining cases, \aspire is significantly more efficient than the flow based importance sampling. These results demonstrate clearly that although importance sampling is desirable for posteriors with significant overlap, the strength of \aspire lies in the way it is still more efficient even with an imperfect or only marginally overlapping proposal, while also being robust against errors propagating from the flow, as by the final resampling stage the flow proposal has no bearing on the posteriors. 
\par
The suite of analyses chosen for this work demonstrates a finite set of applications of this framework for standard siren cosmology. Other use cases to which it could be applied are left for future work, for example expanding the parameter space to include non-$\Lambda$CDM cosmological models, or analysing the most informative events (high SNR, small localisation volume) first and then adding the less well-constrained ones. As gravitational-wave catalogs continue to grow, the cost of reanalysis from the prior each time will become greater. Posterior reuse offers a path for efficient dark siren cosmology in which every result can be a starting point for the next analysis.

\begin{acknowledgements}
A.~Papadopoulos is supported by UKRI STFC studentship 323353-01. M.~J~.~Williams acknowledges partial support from STFC grant UKRI2490. C. E. A.~Chapman-Bird is supported by UKSA grant UKRI971.  R.~Gray and C.~Messenger are supported by United Kingdom Research and Innovation grant UKRI2487. This work makes use of the publicly-available GWTC-5.0 parameter estimation data products parts 1 and 2~\cite{gwtc5_pe_part1,gwtc5_pe_part2}. It also used the associated sensitivity injection campaign~\cite{gwtc5_sensitivity}. This material is based upon work supported by NSF's LIGO Laboratory which is a major facility fully funded by the National Science Foundation. The authors are grateful for computational resources provided by the LIGO Laboratory and supported by National Science Foundation Grants PHY-0757058 and PHY-0823459
\end{acknowledgements}
\section*{Data availability}
The analysis framework used in this work is available at~\cite{papadopoulos_repo_2026}, and can be used with the pip-installable versions of \aspire and \gwcosmo. Any further information required is available by contacting the corresponding author.
\section*{Software}
This work makes use of the following software: \gwcosmo~\cite{gwcosmo2021, gwcosmo2023,Papadopoulos:2026puy}, \aspire~\cite{Williams:2025aar,williams_2026_aspire}, \texttt{ASPIRE-bilby}~\cite{williams_2025_aspire_bilby}, \dynesty~\cite{dynesty_paper,dynesty_v3.1.0}, \texttt{numpy}~\cite{harris2020array}, \texttt{scipy}~\cite{2020SciPy-NMeth}, \texttt{bilby}~\cite{bilby_paper}, \texttt{corner}~\cite{corner}, \texttt{Zuko}~\cite{rozet_2024_zuko}, \texttt{Pytorch}~\cite{paszke2019}, \texttt{numba}~\cite{numba} and \texttt{minipcn}~\cite{williams_2026_minipcn}
\appendix
\section{JS divergence criterion}\label{app:js}
\begin{figure}[htb]
    \includegraphics[width=\linewidth]{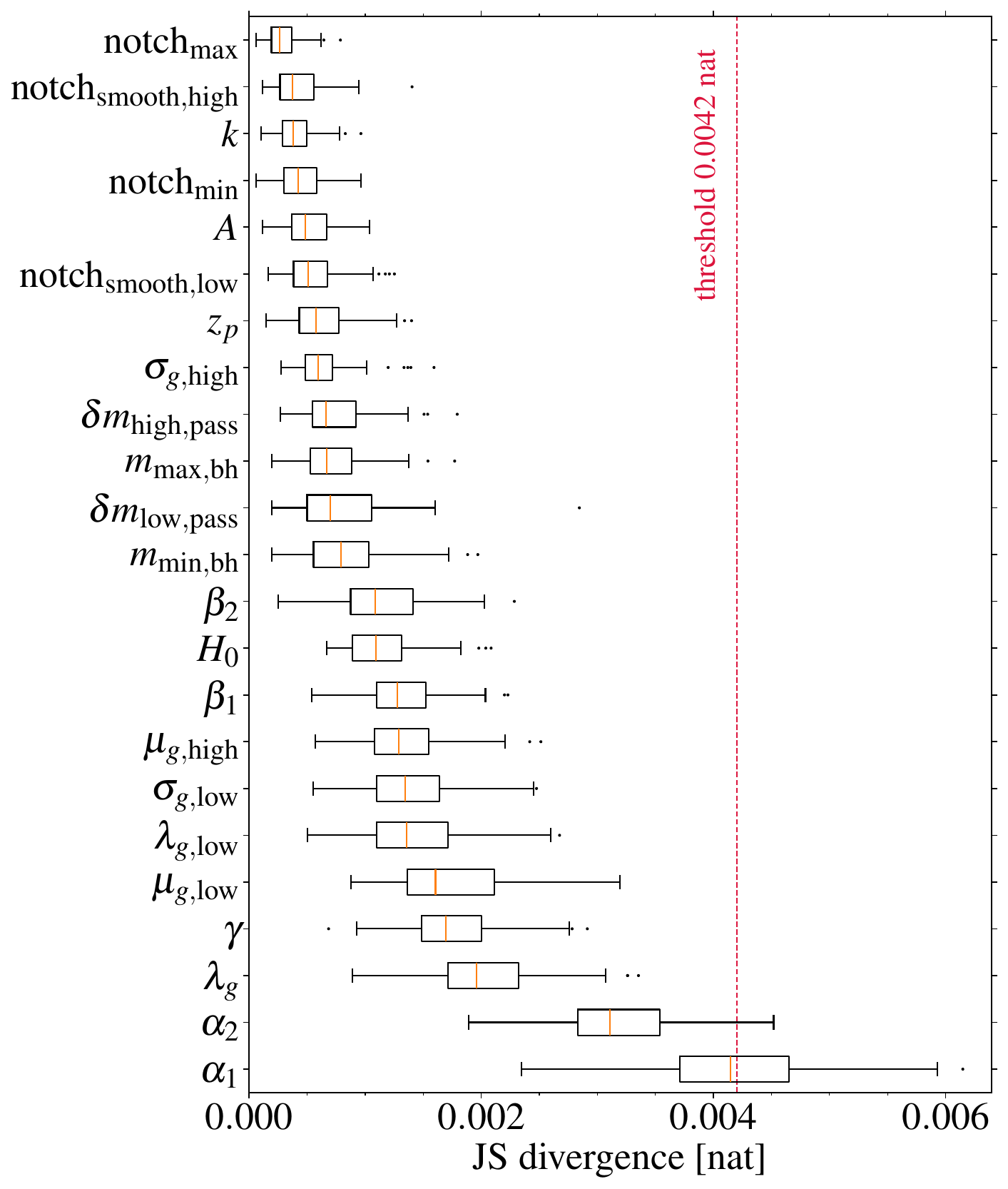}
    \caption{JS divergence distributions for per-parameter marginal posteriors. Values calculated with two identical dark siren analyses using GWTC-5.0, the \fullpop mass model and \textsc{GLADE+}. The boxplots demonstrate the median (in orange) and quartile values of each distribution. The mean of the most divergent parameter ($\alpha_1$), which is used as our threshold, is shown by the red dashed line.}
    \label{fig:js_citerion}
\end{figure}

In order to quantify the agreement between posterior distributions, we follow a similar approach to that used in~\cite{Romero-Shaw:2020owr}. Taking two identical but independent dark siren analyses of GWTC-5.0 using the \fullpop model and the \textsc{GLADE+} catalog with \texttt{dynesty}, we use bootstrapping to create 100 realisations of the posteriors. This was done using a random subsample of 5000 posterior samples from each posterior, to match the expected noise from finite samples with \texttt{ASPIRE} results.
For the marginal posterior of each parameter, the JS divergence is calculated in nat. This is done for each of the 100 realisations of the posterior. Then, to determine the criterion for agreement, the mean JS value of the most divergent parameter is taken. The resulting JS threshold is found to be $0.0042 \,\mathrm{nat}$, or $4.2$ mnat, as can be seen in Fig.~\ref{fig:js_citerion}.
\section{Flow configuration testing}\label{app:flow}
\begin{figure}[htb]
    \includegraphics[width=\linewidth]{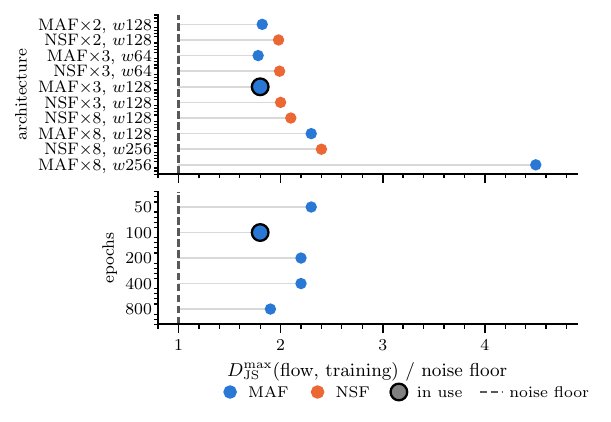}
    \caption{Comparison of different flow configurations as a factor of the JS divergence noise threshold. Different flow types, transform numbers and hidden layer widths are shown in the top panel. The bottom panel takes the best scoring configuration from the top panel (MAF$\times3$, $w128$) and compares it for different training lengths, $N_{\mathrm{ep}}$.}
    \label{fig:flow_test}
\end{figure}

In all analyses presented in this work, we use the default \aspire settings for the \texttt{ZukoFlow} normalising flow. 
To ensure that a poor flow configuration was not degrading our analyses by providing ill-fit proposals, we tested a number of different settings by fitting them to a spectral \gwtc posterior. The resulting flows were sampled and compared to the fitted posterior using the maximum marginal JS divergence. We determined that the base settings using a masked autoregressive flow (MAF) with 3 transforms and a hidden width of 128, trained for 100 epochs provided the better agreement with the true posterior than more complex architectures. Less complex architectures provided a similar level of agreement, but as the cost of the flow is negligible, this gains nothing in terms of overall performance and so we use the base settings. This is shown in Fig.~\ref{fig:flow_test}, and the full set of examined settings are shown in Table~\ref{tab:flow_sweep}.
\begin{table}[htb]
\centering
\small
\setlength{\tabcolsep}{4pt}
\begin{tabular}{@{}lcccccr@{}}
\toprule
Flow & $N_{\mathrm{tr}}$ & $w$ & $N_{\mathrm{ep}}$ & $D_{\mathrm{JS}}$ & $\mathcal{R}$ & $t$ \\
     &                   &     &                   & ($10^{-3}$)       &               & (s) \\
\midrule
MAF & 2 & 128 & 100 & 1.99 & 1.8 & 5 \\
NSF & 2 & 128 & 100 & 2.17 & 2.0 & 6 \\
MAF & 3 & 64  & 100 & 1.96 & 1.8 & 6 \\
NSF & 3 & 64  & 100 & 2.18 & 2.0 & 10 \\
\textbf{MAF} & \textbf{3} & \textbf{128} & \textbf{100} & \textbf{1.97} & \textbf{1.8} & \textbf{8} \\
NSF & 3 & 128 & 100 & 2.16 & 2.0 & 11 \\
NSF & 8 & 128 & 100 & 2.28 & 2.1 & 27 \\
MAF & 8 & 128 & 100 & 2.47 & 2.3 & 7 \\
NSF & 8 & 256 & 100 & 2.61 & 2.4 & 30 \\
MAF & 8 & 256 & 100 & 4.91 & 4.5 & 9 \\
\midrule
MAF & 3 & 128 & 50  & 2.44 & 2.3 & 2 \\
MAF & 3 & 128 & 100 & 1.97 & 1.8 & 8 \\
MAF & 3 & 128 & 200 & 2.39 & 2.2 & 6 \\
MAF & 3 & 128 & 400 & 2.33 & 2.2 & 12 \\
MAF & 3 & 128 & 800 & 2.07 & 1.9 & 23 \\
\bottomrule
\end{tabular}
\caption{Flow configurations tested against a true \gwtc spectral posterior. Columns give the flow type (MAF, masked
autoregressive flow; NSF, neural spline flow), the number of stacked
transforms $N_{\mathrm{tr}}$, the hidden width $w$, the training length $N_{\mathrm{ep}}$, and the fit time $t$ on one H100
GPU. $D_{\mathrm{JS}}$ is the largest marginal Jensen--Shannon divergence over
the $23$ parameters between the flow's own draws and the samples it was
trained on; $\mathcal{R} \equiv D_{\mathrm{JS}}/D_{\mathrm{JS}}^{\mathrm{fl}}$
expresses it in units of the sampling noise floor
$D_{\mathrm{JS}}^{\mathrm{fl}} = 1.08\times10^{-3}$~nat, the same statistic
applied to two independent halves of one spectral posterior, so $\mathcal{R}=1$ is a
perfect fit. This floor is derived separately to the one in \S~\ref{app:js}, as it compares two halves of the samples of one posterior, rather than two independent posteriors. Upper block varies architecture at
fixed $N_{\mathrm{ep}}=100$; lower block varies $N_{\mathrm{ep}}$ at the
architecture in use (bold), which is the best of the ten.}
\label{tab:flow_sweep}

\end{table}
\section{Parameters}\label{app:corner_params}
The corner plots presented in \S~\ref{sec:results} contain a subset of parameters from the cosmological model, the Madau-Dickinson rate evolution model, and the \fullpop mass model. The recovered parameter values from these corner plots and their $68\%$ CI widths are found in Table~\ref{tab:corner_params}. The complete set of parameters highlighted in this work are defined in Table~\ref{tab:params}. This represents a subset of the parameters describing the \fullpop mass mode; for a full description, see~\cite{gwtc4cosmo}.

\begin{table}[t]
\centering\small\setlength{\tabcolsep}{6pt}
\begin{tabular*}{\columnwidth}{@{\extracolsep{\fill}}lcc@{}}
\toprule
 & \aspire & \dynesty \\
\midrule
\multicolumn{3}{@{}l}{\textbf{spectral $\to$ dark}}\\[2pt]
\quad $H_0$       & $73.4^{+14.5}_{-13.6}$  & $73.5^{+14.6}_{-13.8}$  \\
\quad $\gamma$    & $2.75^{+0.58}_{-0.50}$  & $2.77^{+0.58}_{-0.50}$  \\
\quad $m_{\max}$  & $92.6^{+44.8}_{-14.4}$  & $92.2^{+40.5}_{-14.4}$  \\
\quad $\mu_{g,\rm low}$ & $9.02^{+0.36}_{-0.45}$  & $9.02^{+0.35}_{-0.45}$  \\
\addlinespace[3pt]
\multicolumn{3}{@{}l}{\textbf{spectral\textsubscript{\textsc{GWTC-4.0}} $\to$ spectral}}\\[2pt]
\quad $H_0$       & $70.3^{+15.7}_{-14.2}$  & $70.4^{+15.8}_{-13.6}$  \\
\quad $\gamma$    & $2.80^{+0.58}_{-0.49}$  & $2.82^{+0.58}_{-0.52}$  \\
\quad $m_{\max}$  & $92.9^{+38.7}_{-14.1}$  & $92.8^{+41.7}_{-14.2}$  \\
\quad $\mu_{g,\rm low}$ & $9.08^{+0.37}_{-0.52}$  & $9.08^{+0.37}_{-0.53}$  \\
\addlinespace[3pt]
\multicolumn{3}{@{}l}{\textbf{pop $\to$ spectral}}\\[2pt]
\quad $H_0$       & $70.6^{+16.0}_{-14.3}$  & $70.4^{+15.8}_{-13.6}$  \\
\quad $\gamma$    & $2.81^{+0.57}_{-0.52}$  & $2.82^{+0.58}_{-0.52}$  \\
\quad $m_{\max}$  & $92.5^{+37.6}_{-14.3}$  & $92.8^{+41.7}_{-14.2}$  \\
\quad $\mu_{g,\rm low}$ & $9.09^{+0.37}_{-0.52}$  & $9.08^{+0.37}_{-0.53}$  \\
\addlinespace[3pt]
\multicolumn{3}{@{}l}{\textbf{spectral $\to$ pop}}\\[2pt]
\quad $\gamma$    & $2.78^{+0.55}_{-0.50}$  & $2.79^{+0.53}_{-0.53}$  \\
\quad $m_{\max}$  & $91.8^{+42.7}_{-13.0}$  & $92.0^{+40.9}_{-13.2}$  \\
\quad $\mu_{g,\rm low}$ & $9.13^{+0.16}_{-0.27}$  & $9.14^{+0.15}_{-0.24}$  \\
\quad $\mu_{g,\rm high}$ & $28.19^{+1.48}_{-1.67}$ & $28.22^{+1.49}_{-1.53}$ \\
\addlinespace[3pt]
\multicolumn{3}{@{}l}{\textbf{spectral\textsubscript{\mltp} $\to$ spectral}}\\[2pt]
\quad $H_0$       & $70.6^{+15.5}_{-13.9}$  & $70.4^{+15.8}_{-13.6}$  \\
\quad $\gamma$    & $2.79^{+0.59}_{-0.48}$  & $2.82^{+0.58}_{-0.52}$  \\
\quad $m_{\max}$  & $93.3^{+42.7}_{-14.5}$  & $92.8^{+41.7}_{-14.2}$  \\
\quad $\mu_{g,\rm low}$ & $9.07^{+0.37}_{-0.54}$  & $9.08^{+0.37}_{-0.53}$  \\
\bottomrule
\end{tabular*}
\caption{Recovered parameters from posterior distributions for parameters shown in each 
corner plot, comparing the \aspire\ result with its \dynesty\ reference.
Entries are the median with the $68\%$ credible interval. $H_0$ is in
$\mathrm{km\,s^{-1}\,Mpc^{-1}}$, $m_{\max}$, $\mu_{g,\rm low}$ and $\mu_{g,\rm high}$ in
$\mathrm{M_\odot}$, and $\gamma$ is dimensionless; $H_0$ is fixed in the
pop-only analysis, so $\mu_{g,\rm high}$ takes its place there.}
\label{tab:corner_params}
\end{table}

\begin{table}[t]
\centering
\begin{tabular}{@{}llc@{}}
\toprule
Parameter & Description & Prior \\
\midrule
$\Hubble$ & Hubble constant & U(10,200) \\
$\gamma$ & Madau-Dickinson slope index & U($-4$, 12) \\
$m_{\max}$ & Maximum CBC mass [$\mathrm{M_\odot}$] & U(50, 200) \\
$\mu_{g,\rm low}$ & First peak location [$\mathrm{M_\odot}$] & U(5, 150) \\
$\sigma_{g, \rm low}$ & First peak width [$\mathrm{M_\odot}$] & U(0.4, 5) \\
$\mu_{g, \rm high}$ & Second peak location [$\mathrm{M_\odot}$] & U(5, 150) \\
$\lambda_{g}$ & Weight in $\mu_{g,\text{low}}$, $\mu_{g,\rm high}$ & U(0, 1) \\
$\lambda_{g,\rm low}$ & Weight in $\mu_{g,\text{low}}$ against $\mu_{g,\rm high}$ & U(0, 1) \\
\bottomrule
\end{tabular}
\caption{Parameters highlighted in this work, including cosmological parameters, \Hubble, Madau-Dickinson parameters, $\gamma$, and parameters of the \fullpop mass model, $m_{\max}$, $\mu_{g,\rm low}$, $\sigma_{g, \rm low}$, $\mu_{g, \rm high}$, $\lambda_{g}$, $\lambda_{g,\rm low}$. Also included are the prior ranges used for both the \aspire and \dynesty analyses.}\label{tab:params}
\end{table}

\newpage
\bibliography{apssamp}

@article{ligo,
    author = "Aasi, J. and others",
    collaboration = "LIGO Scientific",
    title = "{Advanced LIGO}",
    eprint = "1411.4547",
    archivePrefix = "arXiv",
    primaryClass = "gr-qc",
    doi = "10.1088/0264-9381/32/7/074001",
    journal = "Class. Quant. Grav.",
    volume = "32",
    pages = "074001",
    year = "2015"
}

@article{virgo,
    author = "Acernese, F. and others",
    collaboration = "VIRGO",
    title = "{Advanced Virgo: a second-generation interferometric gravitational wave detector}",
    eprint = "1408.3978",
    archivePrefix = "arXiv",
    primaryClass = "gr-qc",
    doi = "10.1088/0264-9381/32/2/024001",
    journal = "Class. Quant. Grav.",
    volume = "32",
    number = "2",
    pages = "024001",
    year = "2015"
}

@article{kagra,
    author = {Akutsu, T. and others},
    title = {Overview of KAGRA: Detector design and construction history},
    journal = {Progress of Theoretical and Experimental Physics},
    volume = {2021},
    number = {5},
    pages = {05A101},
    year = {2020},
    month = {08},
    issn = {2050-3911},
    doi = {10.1093/ptep/ptaa125},
    url = {https://doi.org/10.1093/ptep/ptaa125},
    eprint = {https://academic.oup.com/ptep/article-pdf/2021/5/05A101/37974994/ptaa125.pdf},
}

@article{schutz1986,
    author = "Schutz, Bernard F.",
    title = "{Determining the Hubble Constant from Gravitational Wave Observations}",
    doi = "10.1038/323310a0",
    journal = "Nature",
    volume = "323",
    pages = "310--311",
    year = "1986"
}

@misc{Williams:2025aar,
    author = "Williams, Michael J.",
    title = "{Accelerated Sequential Posterior Inference via Reuse for Gravitational-Wave Analyses}",
    eprint = "2511.04218",
    archivePrefix = "arXiv",
    primaryClass = "hep-ex",
    reportNumber = "LIGO-P2500623",
    month = "11",
    year = "2025"
}

@misc{Papadopoulos:2026puy,
    author = "Papadopoulos, Alexander and Chapman-Bird, Christian E. A. and Gray, Rachel and Messenger, Christopher and Bertheas, Tom",
    title = "{Scalable Dark Siren Cosmology with gwcosmo: GPU Acceleration, Validation and Systematics}",
    eprint = "2605.23538",
    archivePrefix = "arXiv",
    primaryClass = "astro-ph.CO",
    month = "5",
    year = "2026"
}

@article{Romero-Shaw:2020owr,
    author = "Romero-Shaw, I. M. and others",
    title = "{Bayesian inference for compact binary coalescences with bilby: validation and application to the first LIGO{\textendash}Virgo gravitational-wave transient catalogue}",
    eprint = "2006.00714",
    archivePrefix = "arXiv",
    primaryClass = "astro-ph.IM",
    doi = "10.1093/mnras/staa2850",
    journal = "Mon. Not. Roy. Astron. Soc.",
    volume = "499",
    number = "3",
    pages = "3295--3319",
    year = "2020"
}

@article{gwcosmo2023,
    author = {Gray, Rachel and others},
    year = {2023},
    month = {12},
    pages = {},
    title = {Joint cosmological and gravitational-wave population inference using dark sirens and galaxy catalogues},
    volume = {2023},
    journal = {Journal of Cosmology and Astroparticle Physics},
    doi = {10.1088/1475-7516/2023/12/023}
    }

@article{gwcosmo2021,
    author = {Gray, R and Messenger, C and Veitch, J},
    title = "{A pixelated approach to galaxy catalogue incompleteness: improving the dark siren measurement of the Hubble constant}",
    journal = {Monthly Notices of the Royal Astronomical Society},
    volume = {512},
    number = {1},
    pages = {1127-1140},
    year = {2022},
    month = {02},
    issn = {0035-8711},
    doi = {10.1093/mnras/stac366},
    url = {https://doi.org/10.1093/mnras/stac366},
    eprint = {https://academic.oup.com/mnras/article-pdf/512/1/1127/45303118/stac366.pdf}
}

@article{gwtc4cosmo,
    author = "Abac, A. G. and others",
    collaboration = "LIGO Scientific, Virgo, KAGRA",
    title = "{GWTC-4.0: Constraints on the Cosmic Expansion Rate and Modified Gravitational-wave Propagation}",
    eprint = "2509.04348",
    archivePrefix = "arXiv",
    primaryClass = "astro-ph.CO",
    reportNumber = "LIGO-P2400152",
    doi = "10.3847/2041-8213/ae69dd",
    journal = "Astrophys. J. Lett.",
    volume = "1007",
    number = "1",
    pages = "L17",
    year = "2026"
}

@misc{gwtc5cosmo,
    author = "Abac, A. G. and others",
    collaboration = "LIGO Scientific, VIRGO, KAGRA",
    title = "{GWTC-5.0: Constraints on the Cosmic Expansion Rate and Modified Gravitational-wave Propagation}",
    eprint = "2605.27227",
    archivePrefix = "arXiv",
    primaryClass = "astro-ph.CO",
    reportNumber = "LIGO-P2600018",
    month = "5",
    year = "2026"
}

@misc{gwtc5catalog,
    author = "Abac, A. G. and others",
    collaboration = "LIGO Scientific, VIRGO, KAGRA",
    title = "{GWTC-5.0: Observations from the Second Part of the Fourth LIGO-Virgo-KAGRA Observing Run and Updates to the Gravitational-Wave Transient Catalog}",
    eprint = "2605.27225",
    archivePrefix = "arXiv",
    primaryClass = "gr-qc",
    reportNumber = "LIGO-P2600152",
    month = "5",
    year = "2026"
}

@misc{gwtc4rp,
      title={GWTC-4.0: Population Properties of Merging Compact Binaries}, 
      author={{Abac, A. G.} and others},
        collaboration = {LIGO Scientific, Virgo, and KAGRA Collaborations},
      year={2025},
      eprint={2508.18083},
      archivePrefix={arXiv},
      primaryClass={astro-ph.HE},
      url={https://arxiv.org/abs/2508.18083}, 
}

@article{mddist,
   title={Cosmic Star-Formation History},
   volume={52},
   ISSN={1545-4282},
   url={http://dx.doi.org/10.1146/annurev-astro-081811-125615},
   DOI={10.1146/annurev-astro-081811-125615},
   number={1},
   journal={Annual Review of Astronomy and Astrophysics},
   publisher={Annual Reviews},
   author={Madau, Piero and Dickinson, Mark},
   year={2014},
   month=aug, pages={415–486} }

@misc{gwtc4results,
      title={GWTC-4.0: Updating the Gravitational-Wave Transient Catalog with Observations from the First Part of the Fourth LIGO-Virgo-KAGRA Observing Run}, 
      author={{Abac, A. G.} and others},
      year={2025},
      eprint={2508.18082},
      archivePrefix={arXiv},
      primaryClass={gr-qc},
      url={https://arxiv.org/abs/2508.18082}, }

@article{kldiv,
author = {S. Kullback and R. A. Leibler},
title = {{On Information and Sufficiency}},
volume = {22},
journal = {The Annals of Mathematical Statistics},
number = {1},
publisher = {Institute of Mathematical Statistics},
pages = {79 -- 86},
year = {1951},
doi = {10.1214/aoms/1177729694},
URL = {https://doi.org/10.1214/aoms/1177729694}
}

@article{icarogw2023,
    author = "Mastrogiovanni, Simone and Laghi, Danny and Gray, Rachel and Santoro, Giada Caneva and Ghosh, Archisman and Karathanasis, Christos and Leyde, Konstantin and Steer, Daniele A. and Perries, Stephane and Pierra, Gregoire",
    title = "{Joint population and cosmological properties inference with gravitational waves standard sirens and galaxy surveys}",
    eprint = "2305.10488",
    archivePrefix = "arXiv",
    primaryClass = "astro-ph.CO",
    doi = "10.1103/PhysRevD.108.042002",
    journal = "Phys. Rev. D",
    volume = "108",
    number = "4",
    pages = "042002",
    year = "2023"
}

@article{Palmese_2020,
   title={A Statistical Standard Siren Measurement of the Hubble Constant from the LIGO/Virgo Gravitational Wave Compact Object Merger GW190814 and Dark Energy Survey Galaxies},
   volume={900},
   ISSN={2041-8213},
   url={http://dx.doi.org/10.3847/2041-8213/abaeff},
   DOI={10.3847/2041-8213/abaeff},
   number={2},
   journal={The Astrophysical Journal Letters},
   publisher={American Astronomical Society},
   author={Palmese, A. and others},
   year={2020},
   month=sep, pages={L33} }

@article{GW170817,
   title={A gravitational-wave standard siren measurement of the Hubble constant},
   volume={551},
   ISSN={1476-4687},
   url={http://dx.doi.org/10.1038/nature24471},
   DOI={10.1038/nature24471},
   number={7678},
   journal={Nature},
   publisher={Springer Science and Business Media LLC},
   author={{Abbott, B. P.} and others},
   year={2017},
   month=oct, pages={85–88}
}

@article{bilby_paper,
    author = "Ashton, Gregory and others",
    title = "{BILBY: A user-friendly Bayesian inference library for gravitational-wave astronomy}",
    eprint = "1811.02042",
    archivePrefix = "arXiv",
    primaryClass = "astro-ph.IM",
    doi = "10.3847/1538-4365/ab06fc",
    journal = "Astrophys. J. Suppl.",
    volume = "241",
    number = "2",
    pages = "27",
    year = "2019"
}

@article{holz2005,
    author = "Holz, Daniel E. and Hughes, Scott A.",
    title = "{Using gravitational-wave standard sirens}",
    eprint = "astro-ph/0504616",
    archivePrefix = "arXiv",
    doi = "10.1086/431341",
    journal = "Astrophys. J.",
    volume = "629",
    pages = "15--22",
    year = "2005"
}

@article{moresco2022,
    author = "Moresco, Michele and others",
    title = "{Unveiling the Universe with emerging cosmological probes}",
    eprint = "2201.07241",
    archivePrefix = "arXiv",
    primaryClass = "astro-ph.CO",
    doi = "10.1007/s41114-022-00040-z",
    journal = "Living Rev. Rel.",
    volume = "25",
    number = "1",
    pages = "6",
    year = "2022"
}

@article{chernoff1993,
    author = "Chernoff, David F. and Finn, Lee Samuel",
    title = "{Gravitational radiation, inspiraling binaries, and cosmology}",
    eprint = "gr-qc/9304020",
    archivePrefix = "arXiv",
    reportNumber = "NU-GR-4",
    doi = "10.1086/186898",
    journal = "Astrophys. J. Lett.",
    volume = "411",
    pages = "L5--L8",
    year = "1993"
}

@article{nissanke2010,
    author = "Nissanke, Samaya and Holz, Daniel E. and Hughes, Scott A. and Dalal, Neal and Sievers, Jonathan L.",
    title = "{Exploring short gamma-ray bursts as gravitational-wave standard sirens}",
    eprint = "0904.1017",
    archivePrefix = "arXiv",
    primaryClass = "astro-ph.CO",
    doi = "10.1088/0004-637X/725/1/496",
    journal = "Astrophys. J.",
    volume = "725",
    pages = "496--514",
    year = "2010"
}

@article{taylor2012,
   title={Cosmology using advanced gravitational-wave detectors alone},
   volume={85},
   ISSN={1550-2368},
   url={http://dx.doi.org/10.1103/PhysRevD.85.023535},
   DOI={10.1103/physrevd.85.023535},
   number={2},
   journal={Physical Review D},
   publisher={American Physical Society (APS)},
   author={Taylor, Stephen R. and Gair, Jonathan R. and Mandel, Ilya},
   year={2012},
   month=jan }

@article{ezquiaga2021,
    author = "Ezquiaga, Jose Mar{\'\i}a and Holz, Daniel E.",
    title = "{Jumping the Gap: Searching for LIGO{\textquoteright}s Biggest Black Holes}",
    eprint = "2006.02211",
    archivePrefix = "arXiv",
    primaryClass = "astro-ph.HE",
    doi = "10.3847/2041-8213/abe638",
    journal = "Astrophys. J. Lett.",
    volume = "909",
    number = "2",
    pages = "L23",
    year = "2021"
}

@article{gwtc1,
   title={GWTC-1: A Gravitational-Wave Transient Catalog of Compact Binary Mergers Observed by LIGO and Virgo during the First and Second Observing Runs},
   volume={9},
   ISSN={2160-3308},
   url={http://dx.doi.org/10.1103/PhysRevX.9.031040},
   DOI={10.1103/physrevx.9.031040},
   number={3},
   journal={Physical Review X},
   publisher={American Physical Society (APS)},
   author={Abbott, B. P. and others},
   year={2019},
   month=sep }

@article{gwtc2,
   title={GWTC-2: Compact Binary Coalescences Observed by LIGO and Virgo during the First Half of the Third Observing Run},
   volume={11},
   ISSN={2160-3308},
   url={http://dx.doi.org/10.1103/PhysRevX.11.021053},
   DOI={10.1103/physrevx.11.021053},
   number={2},
   journal={Physical Review X},
   publisher={American Physical Society (APS)},
   author={Abbott, R. and others},
   year={2021},
   month=jun }

@article{gwtc3,
   title={GWTC-3: Compact Binary Coalescences Observed by LIGO and Virgo during the Second Part of the Third Observing Run},
   volume={13},
   ISSN={2160-3308},
   url={http://dx.doi.org/10.1103/PhysRevX.13.041039},
   DOI={10.1103/physrevx.13.041039},
   number={4},
   journal={Physical Review X},
   publisher={American Physical Society (APS)},
   author={Abbott, R. and others},
   year={2023},
   month=dec }

@inbook{vitale2021,
   title={Inferring the Properties of a Population of Compact Binaries in Presence of Selection Effects},
   ISBN={9789811547027},
   url={http://dx.doi.org/10.1007/978-981-15-4702-7_45-1},
   DOI={10.1007/978-981-15-4702-7_45-1},
   booktitle={Handbook of Gravitational Wave Astronomy},
   publisher={Springer Singapore},
   author={Vitale, Salvatore and Gerosa, Davide and Farr, Will M. and Taylor, Stephen R.},
   year={2021},
   pages={1–60} }

@article{mandel2019,
    author = {Mandel, Ilya and Farr, Will M and Gair, Jonathan R},
    title = {Extracting distribution parameters from multiple uncertain observations with selection biases},
    journal = {Monthly Notices of the Royal Astronomical Society},
    volume = {486},
    number = {1},
    pages = {1086-1093},
    year = {2019},
    month = {06},
    issn = {0035-8711},
    doi = {10.1093/mnras/stz896},
    url = {https://doi.org/10.1093/mnras/stz896},
    eprint = {https://academic.oup.com/mnras/article-pdf/486/1/1086/28390969/stz896.pdf},
}

@article{delpozzo2012,
  title = {Inference of cosmological parameters from gravitational waves: Applications to second generation interferometers},
  author = {Del Pozzo, Walter},
  journal = {Phys. Rev. D},
  volume = {86},
  issue = {4},
  pages = {043011},
  numpages = {13},
  year = {2012},
  month = {Aug},
  publisher = {American Physical Society},
  doi = {10.1103/PhysRevD.86.043011},
  url = {https://link.aps.org/doi/10.1103/PhysRevD.86.043011}
}

@article{mastrogiovanni2021,
  title = {On the importance of source population models for gravitational-wave cosmology},
  author = {Mastrogiovanni, S. and Leyde, K. and Karathanasis, C. and Chassande-Mottin, E. and Steer, D. A. and Gair, J. and Ghosh, A. and Gray, R. and Mukherjee, S. and Rinaldi, S.},
  journal = {Phys. Rev. D},
  volume = {104},
  issue = {6},
  pages = {062009},
  numpages = {19},
  year = {2021},
  month = {Sep},
  publisher = {American Physical Society},
  doi = {10.1103/PhysRevD.104.062009},
  url = {https://link.aps.org/doi/10.1103/PhysRevD.104.062009}
}

@article{finke2021,
   title={Cosmology with LIGO/Virgo dark sirens: Hubble parameter and modified gravitational wave propagation},
   volume={2021},
   ISSN={1475-7516},
   url={http://dx.doi.org/10.1088/1475-7516/2021/08/026},
   DOI={10.1088/1475-7516/2021/08/026},
   number={08},
   journal={Journal of Cosmology and Astroparticle Physics},
   publisher={IOP Publishing},
   author={Finke, Andreas and Foffa, Stefano and Iacovelli, Francesco and Maggiore, Michele and Mancarella, Michele},
   year={2021},
   month=aug, pages={026} }

@article{farr2019,
   title={A Future Percent-level Measurement of the Hubble Expansion at Redshift 0.8 with Advanced LIGO},
   volume={883},
   ISSN={2041-8213},
   url={http://dx.doi.org/10.3847/2041-8213/ab4284},
   DOI={10.3847/2041-8213/ab4284},
   number={2},
   journal={The Astrophysical Journal Letters},
   publisher={American Astronomical Society},
   author={Farr, Will M. and Fishbach, Maya and Ye, Jiani and Holz, Daniel E.},
   year={2019},
   month=oct, pages={L42} }

@article{fishbach2018,
   title={Does the Black Hole Merger Rate Evolve with Redshift?},
   volume={863},
   ISSN={2041-8213},
   url={http://dx.doi.org/10.3847/2041-8213/aad800},
   DOI={10.3847/2041-8213/aad800},
   number={2},
   journal={The Astrophysical Journal Letters},
   publisher={American Astronomical Society},
   author={Fishbach, Maya and Holz, Daniel E. and Farr, Will M.},
   year={2018},
   month=Aug, pages={L41} }

@article{borghi2024,
   title={Cosmology and Astrophysics with Standard Sirens and Galaxy Catalogs in View of Future Gravitational Wave Observations},
   volume={964},
   ISSN={1538-4357},
   url={http://dx.doi.org/10.3847/1538-4357/ad20eb},
   DOI={10.3847/1538-4357/ad20eb},
   number={2},
   journal={The Astrophysical Journal},
   publisher={American Astronomical Society},
   author={Borghi, Nicola and Mancarella, Michele and Moresco, Michele and Tagliazucchi, Matteo and Iacovelli, Francesco and Cimatti, Andrea and Maggiore, Michele},
   year={2024},
   month=Mar, pages={191} }

@misc{McMahon:2026nhi,
      title={Measurement of the Hubble constant using the Dark Energy Survey Year 6 Gold galaxy catalog and the fourth Gravitational-Wave Transient Catalog}, 
      author={Isaac McMahon and Danny Laghi and Marcelle Soares-Santos and Kendall Ackley and Gergely Dálya and Yavuz Gençel and David Sánchez-Cid and Felipe Andrade-Oliveira and Sean MacBride and Christian Chapman-Bird and Rachel Gray and Alexander Papadopoulos},
      year={2026},
      eprint={2602.04766},
      archivePrefix={arXiv},
      primaryClass={astro-ph.CO},
      url={https://arxiv.org/abs/2602.04766}, 
}

@article{messenger2012,
   title={Measuring a Cosmological Distance-Redshift Relationship Using Only Gravitational Wave Observations of Binary Neutron Star Coalescences},
   volume={108},
   ISSN={1079-7114},
   url={http://dx.doi.org/10.1103/PhysRevLett.108.091101},
   DOI={10.1103/physrevlett.108.091101},
   number={9},
   journal={Physical Review Letters},
   publisher={American Physical Society (APS)},
   author={Messenger, C. and Read, J.},
   year={2012},
   month=Feb }

@misc{paszke2019,
      title={PyTorch: An Imperative Style, High-Performance Deep Learning Library}, 
      author={Adam Paszke and Sam Gross and Francisco Massa and Adam Lerer and James Bradbury and Gregory Chanan and Trevor Killeen and Zeming Lin and Natalia Gimelshein and Luca Antiga and Alban Desmaison and Andreas Köpf and Edward Yang and Zach DeVito and Martin Raison and Alykhan Tejani and Sasank Chilamkurthy and Benoit Steiner and Lu Fang and Junjie Bai and Soumith Chintala},
      year={2019},
      eprint={1912.01703},
      archivePrefix={arXiv},
      primaryClass={cs.LG},
      url={https://arxiv.org/abs/1912.01703}, 
}

@INPROCEEDINGS{numba,
       author = {{Lam}, Siu Kwan and {Pitrou}, Antoine and {Seibert}, Stanley},
        title = "{Numba: A LLVM-based Python JIT Compiler}",
    booktitle = {Proc. Second Workshop on the LLVM Compiler Infrastructure in HPC},
         year = 2015,
        month = nov,
        pages = {1-6},
          doi = {10.1145/2833157.2833162},
       adsurl = {https://ui.adsabs.harvard.edu/abs/2015llvm.confE...1L}
}

@misc{pierra2026,
      title={Heavy Black-Holes Also Matter in Standard Siren Cosmology}, 
      author={Grégoire Pierra and Alexander Papadopoulos},
      year={2026},
      eprint={2601.03257},
      archivePrefix={arXiv},
      primaryClass={astro-ph.CO},
      url={https://arxiv.org/abs/2601.03257}, 
}

@article{ezquiaga2022,
   title={Spectral Sirens: Cosmology from the Full Mass Distribution of Compact Binaries},
   volume={129},
   ISSN={1079-7114},
   url={http://dx.doi.org/10.1103/PhysRevLett.129.061102},
   DOI={10.1103/physrevlett.129.061102},
   number={6},
   journal={Physical Review Letters},
   publisher={American Physical Society (APS)},
   author={Ezquiaga, Jose María and Holz, Daniel E.},
   year={2022},
   month=Aug }

@article{soaressantos2019,
   title={First Measurement of the Hubble Constant from a Dark Standard Siren using the Dark Energy Survey Galaxies and the LIGO/Virgo Binary–Black-hole Merger GW170814},
   volume={876},
   ISSN={2041-8213},
   url={http://dx.doi.org/10.3847/2041-8213/ab14f1},
   DOI={10.3847/2041-8213/ab14f1},
   number={1},
   journal={The Astrophysical Journal Letters},
   publisher={American Astronomical Society},
   author={{Soares-Santos, M.} and others},
   year={2019},
   month=Apr, pages={L7} }

@misc{bertheas2026,
      title={Spectral sirens cosmology from binary black holes populations with sharper mass features}, 
      author={Tom Bertheas and Vasco Gennari and Danièle Steer and Nicola Tamanini},
      year={2026},
      eprint={2603.06792},
      archivePrefix={arXiv},
      primaryClass={gr-qc},
      url={https://arxiv.org/abs/2603.06792}, 
}

@misc{tagliazucchi2026,
      title={Pushing spectral siren cosmology into the third-generation era: a blinded mock data challenge}, 
      author={Matteo Tagliazucchi and Michele Moresco and Alessandro Agapito and Michele Mancarella and Sarah Ferraiuolo and Simone Mastrogiovanni and Nicola Borghi and Francesco Pannarale and Daniele Bonacorsi},
      year={2026},
      eprint={2602.17756},
      archivePrefix={arXiv},
      primaryClass={astro-ph.CO},
      url={https://arxiv.org/abs/2602.17756}, 
}

@article{cozzumbo2025,
doi = {10.1088/1475-7516/2025/05/021},
url = {https://doi.org/10.1088/1475-7516/2025/05/021},
year = {2025},
month = {may},
publisher = {IOP Publishing},
volume = {2025},
number = {05},
pages = {021},
author = {Cozzumbo, Andrea and Dupletsa, Ulyana and Calderón, Rodrigo and Murgia, Riccardo and Oganesyan, Gor and Branchesi, Marica},
title = {Model-independent cosmology with joint observations of gravitational waves and $\ensuremath{\gamma}$-ray bursts},
journal = {Journal of Cosmology and Astroparticle Physics}
}

@article{mancarella2024,
  title = {Accurate Standard Siren Cosmology with Joint Gravitational-Wave and $\ensuremath{\gamma}$-Ray Burst Observations},
  author = {Mancarella, Michele and Iacovelli, Francesco and Foffa, Stefano and Muttoni, Niccol\`o and Maggiore, Michele},
  journal = {Phys. Rev. Lett.},
  volume = {133},
  issue = {26},
  pages = {261001},
  numpages = {7},
  year = {2024},
  month = {Dec},
  publisher = {American Physical Society},
  doi = {10.1103/PhysRevLett.133.261001},
  url = {https://link.aps.org/doi/10.1103/PhysRevLett.133.261001}
}

@article{Dalya:2021ewn,
    author = "D{\'a}lya, G. and others",
    title = "{GLADE+~: an extended galaxy catalogue for multimessenger searches with advanced gravitational-wave detectors}",
    eprint = "2110.06184",
    archivePrefix = "arXiv",
    primaryClass = "astro-ph.CO",
    doi = "10.1093/mnras/stac1443",
    journal = "Mon. Not. Roy. Astron. Soc.",
    volume = "514",
    number = "1",
    pages = "1403--1411",
    year = "2022"
}

@misc{gray2026,
      title={Expanding the scope of dark siren cosmology: Inferring the population properties of gravitational wave-hosting galaxies}, 
      author={Rachel Gray and Daniel Williams and Alexander Papadopoulos},
      year={2026},
      eprint={2609.04991},
      archivePrefix={arXiv},
      primaryClass={astro-ph.CO},
      url={https://arxiv.org/abs/2609.04991}, 
}

@misc{agapito2026,
      title={Gravitational-wave dark sirens as astrophysical probes: inferring galaxy-merger relations and identifying individual hosts}, 
      author={Alessandro Agapito and Michele Mancarella},
      year={2026},
      eprint={2609.38110},
      archivePrefix={arXiv},
      primaryClass={astro-ph.CO},
      url={https://arxiv.org/abs/2609.38110}, 
}

@INPROCEEDINGS{skilling_proceedings,
       author = {{Skilling}, John},
        title = "{Nested Sampling}",
    booktitle = {Bayesian Inference and Maximum Entropy Methods in Science and Engineering: 24th International Workshop on Bayesian Inference and Maximum Entropy Methods in Science and Engineering},
         year = 2004,
       editor = {{Fischer}, Rainer and {Preuss}, Roland and {Toussaint}, Udo Von},
       series = {American Institute of Physics Conference Series},
       volume = {735},
        month = nov,
    publisher = {AIP},
        pages = {395-405},
          doi = {10.1063/1.1835238},
       adsurl = {https://ui.adsabs.harvard.edu/abs/2004AIPC..735..395S}
}

@article{skilling_paper,
author = {John Skilling},
title = {{Nested sampling for general Bayesian computation}},
volume = {1},
journal = {Bayesian Analysis},
number = {4},
publisher = {International Society for Bayesian Analysis},
pages = {833 -- 859},
year = {2006},
doi = {10.1214/06-BA127},
URL = {https://doi.org/10.1214/06-BA127}
}

@misc{dynesty_v3.1.0,
  author       = {Koposov, Sergey and Speagle, Josh and Barbary, Kyle and
                  Ashton, Gregory and Bennett, Ed and Buchner, Johannes and
                  Scheffler, Carl and Talbot, Colm and Cook, Ben and
                  Guillochon, James and Cubillos, Patricio and
                  Asensio Ramos, Andr{\'e}s and Dartiailh, Matthieu and
                  {Ilya} and Tollerud, Erik and Lang, Dustin and
                  Johnson, Benjamin and {jtmendel} and Higson, Edward and
                  others and {joezuntz}},
  title        = {joshspeagle/dynesty: v3.1.0},
  year         = {2026},
  version      = {v3.1.0},
  publisher    = {Zenodo},
  doi          = {10.5281/zenodo.21417233},
  url          = {https://doi.org/10.5281/zenodo.21417233}
}

@article{dynesty_paper,
       author = {{Speagle}, Joshua S.},
        title = "{DYNESTY: a dynamic nested sampling package for estimating Bayesian posteriors and evidences}",
journal = {Monthly Notices of the Royal Astronomical Society},
         year = {2020},
        month = {apr},
       volume = {493},
       number = {3},
        pages = {3132-3158},
          doi = {10.1093/mnras/staa278},
archivePrefix = {arXiv},
       eprint = {1904.02180},
 primaryClass = {astro-ph.IM},
       adsurl = {https://ui.adsabs.harvard.edu/abs/2020MNRAS.493.3132S}
}

@article{mcmc,
   title={Data Analysis Recipes: Using Markov Chain Monte Carlo*},
   volume={236},
   ISSN={1538-4365},
   url={http://dx.doi.org/10.3847/1538-4365/aab76e},
   DOI={10.3847/1538-4365/aab76e},
   number={1},
   journal={The Astrophysical Journal Supplement Series},
   publisher={American Astronomical Society},
   author={Hogg, David W. and Foreman-Mackey, Daniel},
   year={2018},
   month=May, pages={11} }

@misc{Wolfe:2026dcq,
    author = "Wolfe, Noah E. and Mould, Matthew and Veitch, John and Vitale, Salvatore",
    title = "{Neural Bayesian updates to populations with growing gravitational-wave catalogs}",
    eprint = "2602.20277",
    archivePrefix = "arXiv",
    primaryClass = "astro-ph.IM",
    month = "2",
    year = "2026"
}

@article{corner, doi = {10.21105/joss.00024}, url = {https://doi.org/10.21105/joss.00024}, year = {2016}, publisher = {The Open Journal}, volume = {1}, number = {2}, pages = {24}, author = {Foreman-Mackey, Daniel}, title = {corner.py: Scatterplot matrices in Python}, journal = {Journal of Open Source Software} }

@Article{         harris2020array,
 title         = {Array programming with {NumPy}},
 author        = {Charles R. Harris and K. Jarrod Millman and St{\'{e}}fan J.
                 van der Walt and Ralf Gommers and Pauli Virtanen and David
                 Cournapeau and Eric Wieser and Julian Taylor and Sebastian
                 Berg and Nathaniel J. Smith and Robert Kern and Matti Picus
                 and Stephan Hoyer and Marten H. van Kerkwijk and Matthew
                 Brett and Allan Haldane and Jaime Fern{\'{a}}ndez del
                 R{\'{i}}o and Mark Wiebe and Pearu Peterson and Pierre
                 G{\'{e}}rard-Marchant and Kevin Sheppard and Tyler Reddy and
                 Warren Weckesser and Hameer Abbasi and Christoph Gohlke and
                 Travis E. Oliphant},
 year          = {2020},
 month         = sep,
 journal       = {Nature},
 volume        = {585},
 number        = {7825},
 pages         = {357--362},
 doi           = {10.1038/s41586-020-2649-2},
 publisher     = {Springer Science and Business Media {LLC}},
 url           = {https://doi.org/10.1038/s41586-020-2649-2}
}

@misc{williams_2026_minipcn,
  author       = {Williams, Michael J.},
  title        = {mj-will/minipcn: v0.3.0a1},
  year         = {2026},
  publisher    = {Zenodo},
  howpublished = {Zenodo},
  version      = {v0.3.0a1},
  note         = {Computer software},
  doi          = {10.5281/zenodo.22829770},
  url          = {https://doi.org/10.5281/zenodo.22829770}
}

@misc{williams_2026_aspire,
  author       = {Williams, Michael J. and Garron, Angel and Santoliquido, Filippo and Tissino, Jacopo and Leyde, Konstantin},
  title        = {mj-will/aspire: v0.1.0},
  year         = {2026},
  publisher    = {Zenodo},
  howpublished = {Zenodo},
  version      = {v0.1.0},
  note         = {Computer software},
  doi          = {10.5281/zenodo.21774114},
  url          = {https://doi.org/10.5281/zenodo.21774114}
}

@misc{rozet_2024_zuko,
  author       = {Rozet, Fran{\c{c}}ois and Divo, Felix and Schnake, Simon},
  title        = {probabilists/zuko: Zuko 1.1.0},
  year         = {2024},
  publisher    = {Zenodo},
  howpublished = {Zenodo},
  version      = {1.1.0},
  note         = {Computer software},
  doi          = {10.5281/zenodo.10571785},
  url          = {https://doi.org/10.5281/zenodo.10571785}
}

@ARTICLE{2020SciPy-NMeth,
  author  = {Virtanen, Pauli and Gommers, Ralf and Oliphant, Travis E. and
            Haberland, Matt and Reddy, Tyler and Cournapeau, David and
            Burovski, Evgeni and Peterson, Pearu and Weckesser, Warren and
            Bright, Jonathan and {van der Walt}, St{\'e}fan J. and
            Brett, Matthew and Wilson, Joshua and Millman, K. Jarrod and
            Mayorov, Nikolay and Nelson, Andrew R. J. and Jones, Eric and
            Kern, Robert and Larson, Eric and Carey, C J and
            Polat, {\.I}lhan and Feng, Yu and Moore, Eric W. and
            {VanderPlas}, Jake and Laxalde, Denis and Perktold, Josef and
            Cimrman, Robert and Henriksen, Ian and Quintero, E. A. and
            Harris, Charles R. and Archibald, Anne M. and
            Ribeiro, Ant{\^o}nio H. and Pedregosa, Fabian and
            {van Mulbregt}, Paul and {SciPy 1.0 Contributors}},
  title   = {{{SciPy} 1.0: Fundamental Algorithms for Scientific
            Computing in Python}},
  journal = {Nature Methods},
  year    = {2020},
  volume  = {17},
  pages   = {261--272},
  adsurl  = {https://rdcu.be/b08Wh},
  doi     = {10.1038/s41592-019-0686-2},
}

@misc{fearnhead2010adaptivesequentialmontecarlo,
      title={An Adaptive Sequential Monte Carlo Sampler}, 
      author={Paul Fearnhead and Benjamin M. Taylor},
      year={2010},
      eprint={1005.1193},
      archivePrefix={arXiv},
      primaryClass={stat.CO},
      url={https://arxiv.org/abs/1005.1193}, 
}

@article{DelMoral2012,
  author    = {Del Moral, Pierre and Doucet, Arnaud and Jasra, Ajay},
  title     = {An adaptive sequential {Monte Carlo} method for approximate {Bayesian} computation},
  journal   = {Statistics and Computing},
  year      = {2012},
  month     = sep,
  volume    = {22},
  number    = {5},
  pages     = {1009--1020},
  issn      = {1573-1375},
  doi       = {10.1007/s11222-011-9271-y},
  url       = {https://doi.org/10.1007/s11222-011-9271-y}
}

@article{Papamakarios:2019fms,
    author = "Papamakarios, George and Nalisnick, Eric and Rezende, Danilo Jimenez and Mohamed, Shakir and Lakshminarayanan, Balaji",
    title = "{Normalizing Flows for Probabilistic Modeling and Inference}",
    eprint = "1912.02762",
    archivePrefix = "arXiv",
    primaryClass = "stat.ML",
    doi = "10.5555/3546258.3546315",
    journal = "J. Machine Learning Res.",
    volume = "22",
    number = "1",
    pages = "2617--2680",
    year = "2021"
}

@article{Williams:2025szm,
    author = "Williams, Michael J. and Karamanis, Minas and Luo, Yilin and Seljak, Uro{\v{s}}",
    title = "{Validating sequential Monte Carlo for gravitational-wave inference}",
    eprint = "2506.18977",
    archivePrefix = "arXiv",
    primaryClass = "astro-ph.IM",
    reportNumber = "LIGO-P2500231",
    doi = "10.1093/mnras/staf1458",
    journal = "Mon. Not. Roy. Astron. Soc.",
    volume = "543",
    number = "2",
    pages = "1479--1493",
    year = "2025"
}

@article{Bernardo2011,
author = {Bernardo, Jose and Bayarri, M and Berger, J and Dawid, A and Heckerman, David and Smith, A and West, Mike and Del Moral, Pierre and Doucet, Arnaud and Jasra, Ajay},
year = {2011},
month = {12},
pages = {1-34},
title = {Sequential Monte Carlo for Bayesian Computation},
volume = {8},
journal = {BAYESIAN STATISTICS}
}

@article{Beirnaert_2025,
   title={A Hubble constant estimation with dark standard sirens and galaxy cluster catalogues},
   volume={542},
   ISSN={1365-2966},
   url={http://dx.doi.org/10.1093/mnras/staf1432},
   DOI={10.1093/mnras/staf1432},
   number={4},
   journal={Monthly Notices of the Royal Astronomical Society},
   publisher={Oxford University Press (OUP)},
   author={Beirnaert, Freija and Dálya, Gergely and Ghosh, Archisman},
   year={2025},
   month=Aug, pages={3346–3353} }

@ARTICLE{Phillips1993,
       author = {{Phillips}, M.~M.},
        title = "{The Absolute Magnitudes of Type IA Supernovae}",
   journal={The Astrophysical Journal Letters},
         year = {1993},
        month = {aug},
       volume = {413},
        pages = {L105},
          doi = {10.1086/186970},
       adsurl = {https://ui.adsabs.harvard.edu/abs/1993ApJ...413L.105P}
}

@ARTICLE{Riess1995,
       author = {{Riess}, Adam G. and {Press}, William H. and {Kirshner}, Robert P.},
        title = "{Using Type IA Supernova Light Curve Shapes to Measure the Hubble Constant}",
   journal={The Astrophysical Journal Letters},
         year = 1995,
        month = jan,
       volume = {438},
        pages = {L17},
          doi = {10.1086/187704},
archivePrefix = {arXiv},
       eprint = {astro-ph/9410054},
 primaryClass = {astro-ph},
       adsurl = {https://ui.adsabs.harvard.edu/abs/1995ApJ...438L..17R}
}

@article{hughes2002,
  author  = {{Hughes}, Scott A.},
  title   = {Untangling the merger history of massive black holes with LISA},
  journal = {Monthly Notices of the Royal Astronomical Society},
  year    = {2002},
  volume  = {331},
  pages   = {805-816},
  eprint  = {astro-ph/0108483}
}

@misc{Gennari:2026dfy,
    author = "Gennari, Vasco and Bertheas, Tom and Tamanini, Nicola",
    title = "{Emergent structure in the binary black hole mass distribution and implications for population-based cosmology}",
    eprint = "2604.14290",
    archivePrefix = "arXiv",
    primaryClass = "gr-qc",
    month = "4",
    year = "2026"
}

@misc{Gennari:2025nho,
    author = "Gennari, Vasco and Mastrogiovanni, Simone and Tamanini, Nicola and Marsat, Sylvain and Pierra, Gr{\'e}goire",
    title = "{Searching for additional structure and redshift evolution in the observed binary black hole population with a parametric time-dependent mass distribution}",
    eprint = "2502.20445",
    archivePrefix = "arXiv",
    primaryClass = "gr-qc",
    doi = "10.1103/ftw9-7xd5",
    journal = "Phys. Rev. D",
    volume = "111",
    number = "12",
    pages = "123046",
    year = "2025"
}

@misc{Tagliazucchi:2026gxn,
    author = "Tagliazucchi, Matteo and Moresco, Michele and Borghi, Nicola and Ciapetti, Chiara",
    title = "{Mind the peak: Improving cosmological constraints from GWTC-4.0 spectral sirens using semiparametric mass models}",
    eprint = "2601.03347",
    archivePrefix = "arXiv",
    primaryClass = "astro-ph.CO",
    doi = "10.1051/0004-6361/202558756",
    journal = "Astron. Astrophys.",
    volume = "709",
    pages = "A197",
    year = "2026"
}

@misc{Bertheas:2025mzd,
    author = "Bertheas, Tom and Gennari, Vasco and Tamanini, Nicola",
    title = "{Towards a few percent measurement of the Hubble constant with the current network of gravitational wave detectors without using electromagnetic information}",
    eprint = "2508.19331",
    archivePrefix = "arXiv",
    primaryClass = "gr-qc",
    month = "8",
    year = "2025"
}

@article{Karamanis2022, doi = {10.21105/joss.04634}, url = {https://doi.org/10.21105/joss.04634}, year = {2022}, publisher = {The Open Journal}, volume = {7}, number = {79}, pages = {4634}, author = {Karamanis, Minas and Nabergoj, David and Beutler, Florian and Peacock, John A. and Seljak, Uroš}, title = {pocoMC: A Python package for accelerated Bayesian inference in astronomy and cosmology}, journal = {Journal of Open Source Software} }

@misc{williams_2025_aspire_bilby,
  author    = {Williams, Michael J.},
  title     = {mj-will/aspire-bilby: v0.1.0a2},
  year      = {2025},
  publisher = {Zenodo},
  version   = {v0.1.0a2},
  doi       = {10.5281/zenodo.17540602},
  url       = {https://doi.org/10.5281/zenodo.17540602}
}

@misc{papadopoulos_repo_2026,
  author       = {Papadopoulos, Alexander},
  title        = {aspire-gwcosmo: a framework for sequential standard sire inference},
  year         = {2026},
  howpublished = {\url{https://github.com/alexpapad0poulos/aspire_gwcosmo}}
}

@misc{gwtc5_pe_part1,
  author       = {{LIGO Scientific Collaboration} and {Virgo Collaboration} and {KAGRA Collaboration}},
  title        = {{GWTC-5.0: Parameter estimation data release (Part 1 of 2)}},
  year         = {2026},
  publisher    = {Zenodo},
  howpublished = {Data set},
  doi          = {10.5281/zenodo.20276106},
  url          = {https://doi.org/10.5281/zenodo.20276106}
}

@misc{gwtc5_pe_part2,
  author       = {{LIGO Scientific Collaboration} and {Virgo Collaboration} and {KAGRA Collaboration}},
  title        = {{GWTC-5.0: Parameter estimation data release (Part 2 of 2)}},
  year         = {2026},
  publisher    = {Zenodo},
  howpublished = {Data set},
  doi          = {10.5281/zenodo.20291740},
  url          = {https://doi.org/10.5281/zenodo.20291740}
}

@misc{gwtc5_sensitivity,
  author       = {{LIGO Scientific Collaboration} and {Virgo Collaboration} and {KAGRA Collaboration}},
  title        = {{GWTC-5.0: Cumulative Search Sensitivity Estimates}},
  year         = {2026},
  publisher    = {Zenodo},
  howpublished = {Data set},
  doi          = {10.5281/zenodo.19500052},
  url          = {https://doi.org/10.5281/zenodo.19500052}
}

@misc{gwtc5_methods,
    author = "Abac, A. G. and others",
    collaboration = "LIGO Scientific, VIRGO, KAGRA",
    title = "{GWTC-5.0: Methods for Identifying and Characterizing Gravitational-wave Transients}",
    eprint = "2605.27224",
    archivePrefix = "arXiv",
    primaryClass = "gr-qc",
    reportNumber = "LIGO-P2600166",
    month = "5",
    year = "2026"
}

@article{Essick:2025,
  title = {Compact binary coalescence sensitivity estimates with injection campaigns during the LIGO-Virgo-KAGRA Collaborations' fourth observing run},
  author = {Essick, Reed and Coughlin, Michael W. and Zevin, Michael and Chatterjee, Deep and Clarke, Teagan A. and Colloms, Storm and Mali, Utkarsh and Miller, Simona and Steinle, Nathan and Baral, Pratyusava and Baylor, Amanda C. and Cabourn Davies, Gareth and Dent, Thomas and Joshi, Prathamesh and Kumar, Praveen and Messick, Cody and Mishra, Tanmaya and Ouzriat, Amazigh and Phukon, Khun Sang and Piccari, Lorenzo and Pillas, Marion and Trevor, Max and Callister, Thomas A. and Fishbach, Maya},
  journal = {Phys. Rev. D},
  volume = {112},
  issue = {10},
  pages = {102001},
  numpages = {26},
  year = {2025},
  month = {Nov},
  publisher = {American Physical Society},
  doi = {10.1103/44x3-hv3y},
  url = {https://link.aps.org/doi/10.1103/44x3-hv3y}
}

@misc{Wouters:2026nbh,
    author = "Wouters, Thibeau and Ng, Thomas C. K. and Koehn, Hauke and Gittins, Fabian and Pang, Peter T. H. and Dietrich, Tim and Van Den Broeck, Chris",
    title = "{Scalable and sequential inference of the neutron star equation of state with the Einstein Telescope}",
    eprint = "2610.07975",
    archivePrefix = "arXiv",
    primaryClass = "astro-ph.HE",
    month = "10",
    year = "2026"
}

@article{Essick:2023upv,
    author = "Essick, Reed and Fishbach, Maya",
    title = "{Ensuring Consistency between Noise and Detection in Hierarchical Bayesian Inference}",
    eprint = "2310.02017",
    archivePrefix = "arXiv",
    primaryClass = "gr-qc",
    doi = "10.3847/1538-4357/ad1604",
    journal = "Astrophys. J.",
    volume = "962",
    number = "2",
    pages = "169",
    year = "2024"
}

@misc{DeRenzis:2026hcp,
    author = "De Renzis, Viola and Agapito, Alessandro and Mancarella, Michele",
    title = "{A weakly modelled view of the joint compact-binary mass plane: population structure and spectral-siren cosmology}",
    eprint = "2610.10535",
    archivePrefix = "arXiv",
    primaryClass = "gr-qc",
    month = "10",
    year = "2026"
}

@misc{Leyde:2026bat,
    author = "Leyde, Konstantin and Colangeli, Elena",
    title = "{CosmoPyro: Gradients for Gravitational-Wave Cosmology}",
    eprint = "2608.18281",
    archivePrefix = "arXiv",
    primaryClass = "astro-ph.CO",
    month = "8",
    year = "2026"
}

@article{Li:2026amt,
    author = "Li, Yin-Jie and Wang, Yi-Ying and Wang, Yuan-Zhu and Tang, Shao-Peng and Fan, Yi-Zhong",
    title = "{Secondary-mass Features Improve Spectral-siren H$_{0}$ Constraints}",
    eprint = "2605.11474",
    archivePrefix = "arXiv",
    primaryClass = "astro-ph.CO",
    doi = "10.3847/1538-4357/aea3fc",
    journal = "Astrophys. J.",
    volume = "1010",
    number = "1",
    pages = "81",
    year = "2026"
}

@article{Tong:2025xvd,
    author = "Tong, Hui and Fishbach, Maya and Thrane, Eric",
    title = "{Spinning Spectral Sirens: Robust Cosmological Measurement Using Mass{\textendash}Spin Correlations in the Binary Black Hole Population}",
    eprint = "2502.10780",
    archivePrefix = "arXiv",
    primaryClass = "astro-ph.CO",
    reportNumber = "LIGO Document Number LIGO-P2400105",
    doi = "10.3847/1538-4357/adcec5",
    journal = "Astrophys. J.",
    volume = "985",
    number = "2",
    pages = "220",
    year = "2025"
}

@article{Li:2024rmi,
    author = "Li, Yin-Jie and Tang, Shao-Peng and Wang, Yuan-Zhu and Fan, Yi-Zhong",
    title = "{Multispectral Sirens: Gravitational-wave Cosmology with (Multi-) Subpopulations of Binary Black Holes}",
    eprint = "2406.11607",
    archivePrefix = "arXiv",
    primaryClass = "astro-ph.CO",
    doi = "10.3847/1538-4357/ad888b",
    journal = "Astrophys. J.",
    volume = "976",
    number = "2",
    pages = "153",
    year = "2024"
}

@article{Farah:2024xub,
    author = "Farah, Amanda M. and Callister, Thomas A. and Ezquiaga, Jose Mar{\'\i}a and Zevin, Michael and Holz, Daniel E.",
    title = "{No Need to Know: Toward Astrophysics-free Gravitational-wave Cosmology}",
    eprint = "2404.02210",
    archivePrefix = "arXiv",
    primaryClass = "astro-ph.CO",
    doi = "10.3847/1538-4357/ad9253",
    journal = "Astrophys. J.",
    volume = "978",
    number = "2",
    pages = "153",
    year = "2025"
}

@misc{lacroze20256,
      title={Cosmological Inference with Gravitational-Wave Dark Sirens and 21-cm Intensity Mapping: Forecasts for LIGO-Virgo-KAGRA with MeerKAT and SKAO}, 
      author={Dounia Nanadoumgar-Lacroze and Konstantin Leyde and Steven Cunnington and Tessa Baker and Ian Harrison and Ulyana Dupletsa and Simone Mastrogiovanni and Riccardo Murgia and Tommaso Ronconi and Matteo Schulz and Marta Spinelli},
      year={2026},
      eprint={2610.31081},
      archivePrefix={arXiv},
      primaryClass={astro-ph.CO},
      url={https://arxiv.org/abs/2610.31081}, 
}

\end{document}